\documentclass[11pt]{article}
\usepackage[margin=1in]{geometry} 
\usepackage{amssymb,amsfonts,amsmath,bbm,mathrsfs,stmaryrd}

\usepackage[colorlinks,
             linkcolor=black!50!red,
             pdftitle={Feynman-diagrammatic description of the asymptotics of the time evolution operator in quantum mechanics},
             pdfauthor={Theo Johnson-Freyd},
             pdfsubject={quantum mechanics},
             pdfkeywords={path integral, quantum mechanics},
             pdfproducer={pdfLaTeX},
             pdfpagemode=None,
             bookmarksopen=true
             bookmarksnumbered=true]{hyperref}

\usepackage[amsmath,thmmarks,hyperref]{ntheorem}
\usepackage{cleveref}

\theoremstyle{change}

\newtheorem{defn}[equation]{Definition}
\newtheorem{thm}[equation]{Theorem}

\newtheorem{lemma}[equation]{Lemma}

\theorembodyfont{\upshape}
\theoremsymbol{\ensuremath{\Diamond}}

\theoremstyle{nonumberplain}

\theoremsymbol{\ensuremath{\Box}}

\qedsymbol{\ensuremath{\Box}}

\creflabelformat{equation}{#2(#1)#3} 

\crefname{equation}{equation}{equations}
\crefname{eg}{example}{examples}
\crefname{defn}{definition}{definitions}
\crefname{prop}{proposition}{propositions}
\crefname{thm}{Theorem}{Theorems}
\crefname{lemma}{lemma}{lemmas}
\crefname{cor}{corollary}{corollaries}
\crefname{section}{Section}{Sections}
\crefname{subsection}{Section}{Sections}

\crefformat{equation}{#2equation~(#1)#3} 
\crefformat{eg}{#2example~#1#3} 
\crefformat{defn}{#2definition~#1#3} 
\crefformat{prop}{#2proposition~#1#3} 
\crefformat{thm}{#2Theorem~#1#3} 
\crefformat{lemma}{#2lemma~#1#3} 
\crefformat{cor}{#2corollary~#1#3} 
\crefformat{section}{#2Section~#1#3} 
\crefformat{subsection}{#2Section~#1#3} 

\Crefformat{equation}{#2Equation~(#1)#3} 
\Crefformat{eg}{#2Example~#1#3} 
\Crefformat{defn}{#2Definition~#1#3} 
\Crefformat{prop}{#2Proposition~#1#3} 
\Crefformat{thm}{#2Theorem~#1#3} 
\Crefformat{lemma}{#2Lemma~#1#3} 
\Crefformat{cor}{#2Corollary~#1#3} 
\Crefformat{section}{#2Section~#1#3} 
\Crefformat{subsection}{#2Section~#1#3}

\numberwithin{equation}{subsection}
\newcommand{\Section}[1]{\setcounter{equation}{0}\section{#1}}

\usepackage{tikz}
\usetikzlibrary{arrows,decorations.pathreplacing,decorations.markings,shapes.geometric,through,fit}

\tikzset{
    plaindot/.style={circle,draw,fill,inner sep=1pt},
    dot/.style={rectangle,draw,fill,inner sep=2pt},
    marking/.style={circle,draw,inner sep=1pt},
    Sdot/.style={rectangle,draw,fill,inner sep=2pt},
    twice/.style={double, double distance=1pt},
}

\newcommand\Cc{{\cal C}}
\newcommand\Dd{{\cal D}}
\newcommand\Ff{{\cal F}}
\newcommand\Oo{{\cal O}}
\newcommand\Pp{{\cal P}}
\newcommand\Qq{{\cal Q}}

\newcommand\CC{{\mathbb C}}
\newcommand\RR{{\mathbb R}}

\newcommand\T{{\rm T}}

\newcommand\st{{\textrm{ s.t.\ }}}
\newcommand\pt{\{{\rm pt}\}}

\newcommand\<{\langle}
\renewcommand\>{\rangle}
\newcommand\tofrom{\leftrightarrow}
\newcommand\sminus{\smallsetminus}

\newcommand\project{\pi}

\DeclareMathOperator{\Aut}{Aut}
\newcommand{\SE}{\hat{H}}
\newcommand{\formal}{\text{\rm formal}}

\newcommand{\define}[1]{{\bf #1}}

\title{Feynman-diagrammatic description of the asymptotics of the time evolution operator in quantum mechanics}
\author{Theo Johnson-Freyd}

\begin{document}
\maketitle
\begin{abstract}
  We describe the ``Feynman diagram'' approach to nonrelativistic quantum mechanics on $\RR^n$, with magnetic and potential terms.  In particular, for each classical path $\gamma$ connecting points $q_0$ and $q_1$ in time $t$, we define a formal power series $V_\gamma(t,q_0,q_1)$ in $\hbar$, given combinatorially by a sum of diagrams that each represent finite-dimensional convergent integrals.  We prove that $\exp(V_\gamma)$ satisfies Schr\"odinger's equation, and explain in what sense the $t\to 0$ limit approaches the $\delta$ distribution.  As such, our construction gives explicitly the full $\hbar\to 0$ asymptotics of the fundamental solution to Schr\"odinger's equation in terms of solutions to the corresponding classical system.  These results justify the heuristic expansion of Feynman's path integral in diagrams.
  
  \noindent\emph{Key words:} quantum mechanics, Feynman diagrams, formal integrals, path integrals, semiclassical asymptotics \\
  \emph{MSC2010:} 81T18, 81S40, 81Q15
\end{abstract}

\Section{Introduction}

The goal of this paper is to show that the semiclassical description of the evolution operator in quantum mechanics in terms of Feynman diagrams satisfies all the necessary requirements: it satisfies the Sch\"odinger equation and has the correct initial conditions.
This description is rooted in Feynman's path integral interpretation of quantum mechanics, first published in \cite{Feynman1948} (see also \cite{FeynmanHibbs1965}), and formal semiclassical arguments.  In certain cases, non-perturbative path integrals can be interpreted analytically as Weiner integrals, e.g.\ \cite {AHK1977,JL2000,KlauderDaubechies1984}.
However, for the last half century, physicists have used ``formal'' perturbative path integrals and Feynman diagrams extensively to make perturbative empirical predictions in theories that do not have adequate mathematical foundations, and there have been similar mathematical applications to the subject of topological quantum field theory \cite{KolyaLong}.  Surprisingly, the diagrammatic methods have not previously been shown to give the correct answers when applied to the one mathematically rigorous nonperturbative theory available, namely nonrelativistic quantum mechanics.  We fill that gap.

In this paper we consider quantum mechanics in $\RR^n$ with the Hamiltonian being a second-order differential operator.  
(In a \cite{melong} we discuss perturbative quantum mechanics on general Riemannian manifolds.)
Recall some basic definitions.
Let $B_1,\dots, B_n, C : \RR^n \to \RR$ be smooth functions.\footnote{Without much difficulty, these functions can be taken to depend explicitly on an external time parameter, but for notational convenience we do not do so.  The function $U(t,q_0,q_1)$ that we will discuss should then be a function of two time variables $U(t_0,t_1,q_0,q_1)$, and similar superficial changes must be made throughout this paper.}  The nonrelativistic Schr\"odinger operator with electric potential $C$ and magnetic potential $B = (B_1,\dots,B_n)$ is the following linear second-order differential operator on $\RR^n$:
\begin{equation} \label{SE}
  \SE = \sum_{j=1}^n \left( i\hbar \frac{\partial}{\partial q^j} + B_j(q)\right)^2 + C(q)
\end{equation}
Solutions to the  partial differential equation $\frac{\partial}{\partial t}\psi(t,q) = \frac1{i\hbar}\SE\psi(t,q)$ with fixed initial data
define a one-parameter family of unitary ``time evolution'' operators $U(t)$ on $L^2(\RR^n)$.  The kernel of the time evolution operator is a distribution $U(t,q_0,q_1)$, satisfying $(U(t)\psi)(q_1) = \int_{\RR^n} U(t,q_0,q_1)\psi(q_0) dq_0$.  It is determined by the following initial value problem:
\begin{align}\label{IVPforU}
  i\hbar \frac{\partial}{\partial t}U(t,q_0,q_1) & = \SE_{q_1}\bigl[U(t,q_0,q_1)\bigr] && U(0,q_0,q) = \delta(q - q_0)
\end{align}
The operator $\SE_{q_1}$ acts on the $q_1$ variable, leaving $(t,q_0)$ fixed.
In this paper we describe explicitly the asymptotics or $U(t,q_0,q_1)$ as $\hbar\to 0$, at least for $(t,q_0,q_1)$ in a dense open subset of $\RR_{>0} \times \RR^{2n}$.

The asymptotics of $U(t,q_0,q_1)$ are known to be given by the Hamilton-Jacobi action and solutions to the transport equation \cite{Takhtajan2008}.  Feynman's path integral also predicts that the asymptotics are given by a power series 
where coefficients are parameterized by
``Feynman diagrams'': each diagram represents a finite-dimensional integral, and the coefficients of the power series are the values of these intergals.  In this paper we will define this diagrammatic expansion; the integrands depend on the solutions to the corresponding classical mechanical system. 
We will prove that the Feynman-diagrammatic power series satisfies formal versions of \cref{IVPforU}: it satisfies the Schr\"odinger equation as a function valued in formal power series, and as a distribution valued in formal power series approaches the $\delta$ distribution as $t\to 0$.

In select cases these results are already known.  For the quadratic ``harmonic oscillator'' (the functions $B_1,\dots,B_n$ are linear in position, $C$ is quadratic), the path integral is defined directly from Gauss's formula, and agrees with the Schr\"odinger picture \cite{Takhtajan2008}.  Feynman's original argument applies when the corresponding classical physics is a perturbation of the free theory, i.e.\ when $B_1,\dots,B_n,C$ are infinitesimal quantities.  Combining these observations, it is not too difficult to prove that the diagrammatic path integral satisfies Schr\"odinger's equation for infinitesimal perturbations of the harmonic oscillator.  More importantly, Duru and Kleinert \cite{DKlong} apply the path integral approach to calculate the energy spectrum of a particle moving in the Coulomb potential, and compare their approach to the Schr\"odinger methods.  Their methods differ from ours: we disallow singular potentials but consider arbitrary smooth ones, so our results do not apply to the Coulomb potential; moreover, they take advantage of numerous changes of variables (without explicitly checking the coordinate invariance of the path integral) and use the ``charge'' of the potential well as the perturbation parameter, whereas we define the path integral directly and use Planck's constant $\hbar$ for perturbation.  Diagrammatic path integrals do exist in the work of Kleinert and Chervyakov \cite{KC5} and in the work of DeWitt-Morette \cite{DWM1976}; Kleinert and Chervyakov's methods apply directly only to perturbations of free motion, and neither they nor DeWitt-Morette (who does give an expansion equivalent to ours) check the Schr\"odinger equation directly, although the result is implicit in their and Duru and Kleinert's approaches.

We summarize the main definitions and results in \cref{OverviewSection}.  \Cref{lemmatasection} has proofs of two key lemmas.  The main theorem is proven in \cref{MainThmProof}, and in \cref{IVPthmproof} we prove the short-time asymptotics.

\subsection{Acknowledgements}

This project was suggested by N.\ Reshetikhin, who provided support and suggestions throughout all stages of it.  K.\ Datchev, C.\ Schommer-Pries, G.\ Thompson, and I.\ Ventura provided valuable discussions.  I would like to also thank the anonymous referee for alerting me to the work of H. Kleinert and collaborators.  I am grateful to Aarhus University for the hospitality.  This work is supported by NSF grant DMS-0901431.

\Section{Statements of the definitions and results} \label{OverviewSection}

\subsection{Preliminaries}

We begin by establishing some notation.  Kronecker's $\delta$-matrix is $\delta^i_j = 1$ if $i=j$ and $0$ otherwise.  Dirac's $\delta$-function is the distribution defined by $\int_{-\infty}^\infty f(x)\delta(x)dx = f(0)$.  Heaviside's step function is:
\[
  \Theta(x) = \int_{-\infty}^x \delta(t)\,dt = \begin{cases} 0, & x < 0 \\ \frac12, & x = 0 \\ 1, & x > 0 \end{cases}
\]
We henceforth adopt Einstein's summation convention: $p_iq^i = q^ip_i = \sum_{i=1}^n p_iq^i$.  We raise and lower indices as is convenient: $p^i = p_i$.

\begin{defn}
  A \define{(piecewise-smooth parameterized) path} is a continuous function $\gamma: [0,t] \to \RR^n$ such that there exists a decomposition $0 = t_0 < t_1 < \dots < t_n = t$ with $\gamma|_{[t_j,t_{j+1}]}$ smooth for each $j=0,\dots,n-1$.  The \define{duration} of a path $\gamma:[0,t] \to \RR^n$ is the real number $t$.  A \define{based loop} is a path $\gamma : [0,t] \to \RR^n$ with $\gamma(0) = 0 = \gamma(t)$.  A path $\gamma$ is \define{classical} if it satisfies the following nonlinear second-order differential equation (we write $\dot\gamma(\tau) = \frac{d}{d\tau}\gamma(\tau)$, etc.):
  \begin{equation} \label{EOM}
    0 = \ddot\gamma^i(\tau) + \left( \frac{\partial B_i}{\partial q^j}\bigl(\gamma(\tau)\bigr) - \frac{\partial B_j}{\partial q^i}\bigl(\gamma(\tau)\bigr) \right) \dot\gamma^j(\tau) + \frac{\partial C}{\partial q^i}\bigl(\gamma(\tau)\bigr)
  \end{equation}
  Let $\gamma,\xi$ be paths of duration $t$, and define the differential operator $\Dd_\gamma$ given by:
  \begin{multline} \label{Dequation}
    \Dd_\gamma[\xi]^i(\tau) = \ddot\xi^i(\tau) + \left( \frac{\partial B_i}{\partial q^j}\bigl(\gamma(\tau)\bigr) - \frac{\partial B_j}{\partial q^i}\bigl(\gamma(\tau)\bigr) \right)\dot\xi^j(\tau) +\\ + \left( \frac{\partial^2 B_i}{\partial q^j\partial q^k}\bigl(\gamma(\tau)\bigr)\,\dot\gamma^k(\tau) - \frac{\partial^2 B_j}{\partial q^i\partial q^k}\bigl(\gamma(\tau)\bigr)\,\dot\gamma^k(\tau) + \frac{\partial^2 C}{\partial q^i\partial q^j}\bigl(\gamma(\tau)\bigr) \right) \xi^j(\tau)
  \end{multline}
  A classical path $\gamma$ of duration $t$ is \define{nondegenerate} if the kernel of $\Dd_\gamma$ among based loops of duration $t$ is trivial.
\end{defn}

As \cref{EOM} is nondegenerate second-order, a classical path $\gamma$ is determined by its duration and its initial conditions $(\dot\gamma(0),\gamma(0))$.  Thus the space of classical paths is naturally an open subset of $\RR^{2n+1}$.  Nondegeneracy is clearly an open condition among classical paths.  A standard result holds that for given initial conditions, the set of durations for which a classical path fails to be nondegenerate is discrete; see e.g.\ \cite{Milnor1963}.  In particular, the nondegenerate classical paths are a dense open subset of all classical paths.  The following lemma follows from e.g.\ \cite{Chris2009}; in \cref{ChrisLemmaProof} we include the proof rewritten in the language of classical paths.

\begin{lemma} \label{ChrisLemma}
  Let $\gamma: [0,t] \to \RR^n$ be classical and nondegenerate.  Then there exists a neighborhood $\Oo$ of $(\gamma(0),\gamma(t)) \in \RR^{2n}$ and a map $\hat\gamma :[0,t] \times \Oo \to \RR^n$ such that (i) $\hat\gamma(-,q_0,q_1)$ is classical and nondegenerate for each $(q_0,q_1) \in \Oo$, (ii) $\hat\gamma(0,q_0,q_1) = q_0$ and $\hat\gamma(t,q_0,q_1) = q_1$, and (iii) $\hat\gamma(-,\gamma(0),\gamma(t)) = \gamma$.
\end{lemma}

Let $\gamma$ be a classical path of duration $t$.  Then for sufficiently small $\epsilon$, $\gamma$ determines classical paths $\gamma_s$ for $s\in [t- \epsilon,t+\epsilon]$, either by restriction or by extension via \cref{EOM}, and if $\gamma$ is nondegenerate, for sufficiently small $\epsilon$ so are the $\gamma_s$.  Thus, to a nondegenerate classical path $\gamma$ we can associate a family $\hat{\gamma_t}(q_0,q_1,-)$ of nondegenerate classical paths, varying smoothly in the parameters.  We will abuse notation and write $\gamma$ for this whole family.  So henceforth $\gamma$ is a smooth family of solutions to the boundary value problem given by \cref{EOM} along with Dirichlet boundary values $\gamma(0) = q_0$, $\gamma(t) = q_1$, where $(t,q_0,q_1)$ ranges over an open subset $\Oo$ of $\RR^{2n+1}$.

\begin{defn}
  Let $\gamma$ be a classical nondegenerate path, extended to a family with boundary values varying in $\Oo\subseteq \RR^{2n+1}$ as above.  The corresponding \define{Hamilton function} $S_\gamma:\Oo \to \RR$ is:
  \begin{equation} \label{Sdefine}
    S_\gamma(t,q_0,q_1) = \int_0^t \left( \frac12 \bigl| \dot\gamma(\tau)\bigr|^2 + B_i\bigl(\gamma(\tau)\bigr)\,\dot\gamma^i(\tau) - C\bigl(\gamma(\tau)\bigr) \right)d\tau
  \end{equation}
\end{defn}
Then $S_\gamma(t,q_0,q_1)$ satisfies the following well-known equations:
\begin{gather} \label{wellknown1}
  \frac{\partial [-S_\gamma]}{\partial t} 
   = \frac12  \bigl| \nabla_{q_0}[-S_\gamma] - B(q_0)\bigr|^2 + C(q_0)
   = \frac12  \bigl| \nabla_{q_1}[-S_\gamma] + B(q_1)\bigr|^2 + C(q_1) 
  \\ \label{wellknown2}
  \frac{\partial [-S_\gamma]}{\partial q_0^i} = \dot\gamma^i(0) + B_i(q_0)
  \quad\quad\quad\quad
  \frac{\partial S_\gamma}{\partial q_1^i} = \dot\gamma^i(t) + B_i(q_1)
\end{gather}

For a few different constructions, we will need the matrix $\frac{\partial^2}{\partial q_0\partial q_1}\bigl[-S(t,q_0,q_1)\bigr]$ and its inverse; we define the indices on the inverse matrix by: $\Bigl( \! \bigl( \frac{\partial^2 [-S]}{\partial q_1\partial q_0}\bigr)^{\!-1}\Bigr)^{\!kl} \frac{\partial^2[-S]}{\partial q_0^l \partial q_1^m} = \delta^k_m$, and similarly for $q_1 \tofrom q_0$.

\begin{lemma} \label{Glemma}
  Let $\gamma$ be a nondegenerate path of duration $t$, extended to a family as above.  Let $G_\gamma: [0,t]^{\times 2} \to (\RR^n)^{\otimes 2}$ be defined by:
  \begin{multline} \label{Gdefine}
    G_\gamma^{ij}(\varsigma,\tau) = \Theta(\varsigma - \tau) \, \frac{\partial \gamma^i}{\partial q_0^k}(\varsigma) \biggl( \! \Bigl( \frac{\partial^2 (-S_\gamma)}{\partial q_0\partial q_1}\Bigr)^{\!-1}\biggr)^{\!kl} \frac{\partial \gamma^j}{\partial q_1^l}(\tau) + \mbox{}
    \\
    \mbox{} + \Theta(\tau - \varsigma)\, \frac{\partial \gamma^i}{\partial q_1^k}(\varsigma) \biggl( \! \Bigl( \frac{\partial^2 (-S_\gamma)}{\partial q_1\partial q_0}\Bigr)^{\!-1}\biggr)^{\!kl} \frac{\partial \gamma^j}{\partial q_0^l}(\tau)
  \end{multline}
  Then $-G_\gamma$ is a \define{Green's function} for the operator $\Dd_\gamma$ given in \cref{Dequation}: i.e.\ $G_\gamma$ satisfies $\Dd_\gamma[G_\gamma]^i_k(\varsigma,\tau) = -\delta^i_k\delta(\tau-\varsigma)$, where $\Dd_\gamma$ acts on the $j,\tau$ parts of \cref{Gdefine}, and $G_\gamma$ vanishes along the boundary of the square $[0,t]^{\times 2}$.
\end{lemma}
For the proof, see \cref{Gproof}.  Statements similar to \cref{ChrisLemma,Glemma} appear in e.g.\ \cite{DWM1976}, where degenerate paths are also briefly addressed.

\subsection{Feynman diagrams and the main results}

\begin{defn}
  A \define{partition} of a set $X$ is a collection of nonempty subsets of $X$, called the \define{blocks} of the partition, that are pairwise disjoint and whose union is all of $X$.  
  
  A \define{Feynman diagram} is a finite combinatorial graph with all vertices trivalent and higher.  (Self-loops, parallel edges, and the empty diagram are allowed.) More precisely, a Feynman diagram is a finite set $H$ of ``half-edges'', along with two partitions $E$ (the ``edges'') and $V$ (the ``vertices'') of $H$, with the requirements $|e| = 2$ for each block $e\in E$, and $|v| \geq 3$ for each block $v\in V$.    An \define{isomorphism} of Feynman diagrams $\Gamma_1 = (H_1,E_1,V_1)$ and $\Gamma_2 = (H_2,E_2,V_2)$ is a bijection $\phi: H_1 \overset\sim\to H_2$ such that $\phi(E_1) = E_2$ and $\phi(V_1) = V_2$, where $\phi(E_1), \phi(V_1)$ are the obvious partitions of $H_2$ induced by $(\phi,E_1,V_1)$. We will denote Feynman diagrams with the obvious pictures:
  
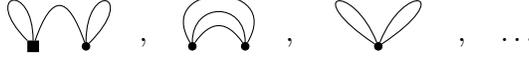
\[
  \begin{tikzpicture}[baseline=(X)]
    \path node[plaindot] (O) {} ++(0pt,2pt) coordinate (X);
    \draw (O) .. controls +(20pt,20pt) and +(0pt,25pt) .. (O);
    \path (O) ++(-20pt,0pt) node[dot] (O1) {};
    \draw (O1) .. controls +(-20pt,20pt) and +(0pt,25pt) .. (O1);
    \draw (O) .. controls +(-7pt,20pt) and +(7pt,20pt) .. (O1);
  \end{tikzpicture}  
  ,
  \begin{tikzpicture}[baseline=(X)]
    \path node[plaindot] (O) {} ++(0pt,2pt) coordinate (X);
    \path (O) ++(-20pt,0pt) node[plaindot] (O1) {};
    \draw (O) .. controls +(-7pt,10pt) and +(7pt,10pt) .. (O1);
    \draw (O) .. controls +(0pt,17pt) and +(0pt,17pt) .. (O1);
    \draw (O) .. controls +(15pt,25pt) and +(-15pt,25pt) .. (O1);
  \end{tikzpicture}  
  ,
  \begin{tikzpicture}[baseline=(X)]
    \path node[plaindot] (O) {} ++(0pt,2pt) coordinate (X);
    \draw (O) .. controls +(30pt,20pt) and +(10pt,25pt) .. (O);
    \draw (O) .. controls +(-30pt,20pt) and +(-10pt,25pt) .. (O);
  \end{tikzpicture}  
  , \quad \dots
  \]
  
  Let $\Gamma = (H,E,V)$ be a Feynman diagram.  The \define{Euler characteristic} of $\Gamma$ is $\chi(\Gamma) = |V| - |E|$.  The \define{connected components of $\Gamma$} are the equivalence classes of $H$ under the equivalence relation generated by $E\cup V$.  Let $\pi_0(\Gamma)$ be the set of connected components of $\Gamma$; then $\lambda(\Gamma) = |\pi_0(\Gamma)| - \chi(\Gamma)$ is the \define{number of loops of $\Gamma$}, also called its \define{first Betti number}.   A diagram $\Gamma$ is \define{connected} if it has precisely one connected component.  For example, the above three pictures comprise all (up to isomorphism) diagrams $\Gamma$ with $\chi(\Gamma) = -1$, and each is connected.
  
  A \define{marked Feynman diagram} is a Feynman diagram $\Gamma = (H,E,V)$ along with a subset $M \subseteq H$ of ``marked half-edges'' such at each vertex $v\in V$, at most one half edge $\eta\in v$ is marked, i.e.\ $|v\cap M|\leq 1$.  An isomorphism of marked Feynman diagrams is required to respect the marking.  We indicate the markings on a marked Feynman diagram with small circles, as in \cref{FeynmanRulesB}.
\end{defn}

\begin{defn}
  Pick a duration $t>0$.  A \define{totally labeled Feynman diagram} (with duration $t$) is a Feynman diagram $\Gamma = (H,E,V)$ along with maps $\vec\tau: V\to [0,t]$ and $\vec\jmath: H \to \{1,\dots,n\}$.  Let $\gamma$ be a classical nondegenerate path of duration $t$ and $(\Gamma,M,\vec\tau,\vec\jmath)$ a totally labeled marked Feynman diagram.  Then its \define{value} is (no summation convention):
  \begin{multline*}
    \Ff_\gamma(\Gamma,M,\vec\tau,\vec\jmath) = \prod_{{ v \in V,\, v\cap M = \emptyset}} \left( 
    \left.\left(\prod_{\eta \in V} \frac{\partial}{\partial q^{\vec\jmath(\eta)}} \right)\bigl[C(q)\bigr] \right|_{q = \gamma(\vec\tau(v))} 
    +
    \left.\left(\prod_{\eta \in V} \frac{\partial}{\partial q^{\vec\jmath(\eta)}} \right)\bigl[B_k(q)\bigr] \right|_{q = \gamma(\vec\tau(v))} \dot\gamma^k(\tau)
    \right) \times \\
    \times \prod_{{ v \in V,\, v\cap M = \{\zeta\}}} \left( \left.\left(\prod_{\eta \in v\sminus \{\zeta\}} \frac{\partial}{\partial q^{\vec\jmath(\eta)}} \right)\bigl[- B_{\vec\jmath(\zeta)}(q)\bigr] \right|_{q = \gamma(\vec\tau(v))} \right) \times \\
    \times \prod_{\substack{e = \{\eta,\zeta\} \in E, \\ e\cap M = \emptyset}} G^{\vec\jmath(\eta),\vec\jmath(\zeta)}(\vec\tau({\eta}),\vec\tau({\zeta}))
     \times \prod_{\substack{e = \{\eta,\zeta\} \in E, \\ e\cap M = \{\eta\}}} \left.\frac{\partial G^{\vec\jmath(\eta),\vec\jmath(\zeta)}(\varsigma_1,\varsigma_2)}{\partial \varsigma_1} \right|_{\substack{ \varsigma_1 = \vec\tau(\eta) \\ \varsigma_2 = \vec\tau(\zeta)}} 
     \times \prod_{\substack{e = \{\eta,\zeta\} \in E, \\ e\subseteq M}} \left.\frac{\partial^2 G^{\vec\jmath(\eta),\vec\jmath(\zeta)}(\varsigma_1,\varsigma_2)}{\partial \varsigma_1\partial \varsigma_2} \right|_{\substack{ \varsigma_1 = \vec\tau(\eta) \\ \varsigma_2 = \vec\tau(\zeta)}} 
  \end{multline*}
If $(\Gamma,M)$ is an (unlabeled) marked Feynman diagram, its \define{value} is an integral over all total labelings:
 \[\Ff_\gamma(\Gamma,M) =\sum_{\vec\jmath\in \{1,\dots,n\}^H} \; \int_{\vec\tau \in [0,t]^V} \limits \Ff_\gamma(\Gamma,M,\vec\tau,\vec\jmath)\,d\vec\tau\]

More compactly, Feynman diagrams are evaluated via the following \define{Feynman rules} and the obvious compositions and contractions (including the Einstein summation convention), where $\xi_1,\dots,\xi_k$ are test-functions ranging over the space of paths of duration $t$:
\begin{gather} \label{FeynmanRulesC}
   \Ff_\gamma\biggl( \begin{tikzpicture}[baseline=(X)]
    \path node[plaindot] (O) {} ++(0pt,4pt) coordinate (X);
    \draw (O) -- ++(-15pt,15pt) +(0,3pt) node[anchor=base] {$\scriptstyle \xi_1$};
    \draw (O) -- ++(-6pt,15pt) +(0,3pt) node[anchor=base] {$\scriptstyle \xi_2$};
    \draw (O) -- ++(15pt,15pt) +(0,3pt) node[anchor=base] {$\scriptstyle \xi_n$};
    \path (O) ++(4pt,13pt) node {$\scriptstyle \ldots$};
  \end{tikzpicture}  \biggr)
   = \int_{\tau = 0}^t \left.\left(
   \frac{\partial^n C(q)}{\partial q^{i_1}\cdots \partial q^{i_n}}
   +
   \frac{\partial^n B_k(q)}{\partial q^{i_1}\cdots \partial q^{i_n}}\, \dot\gamma^k(\tau)
   \right) \right|_{q = \gamma(\tau)}  \xi_1^{i_1}(\tau)\, \cdots\, \xi_n^{i_n}(\tau)\,d\tau \\
 \label{FeynmanRulesB}
   \Ff_\gamma\biggl( \begin{tikzpicture}[baseline=(X)]
    \path node[plaindot] (O) {} ++(0pt,4pt) coordinate (X);
    \draw (O) -- node[marking] {} ++(-15pt,15pt) +(0,3pt) node[anchor=base] {$\scriptstyle \xi_1$};
    \draw (O) -- ++(-6pt,15pt) +(0,3pt) node[anchor=base] {$\scriptstyle \xi_2$};
    \draw (O) -- ++(15pt,15pt) +(0,3pt) node[anchor=base] {$\scriptstyle \xi_n$};
    \path (O) ++(4pt,13pt) node {$\scriptstyle \ldots$};
  \end{tikzpicture}  \biggr)
   = \int_{\tau = 0}^t \left.\frac{\partial^{n-1} [-B_{i_1}]}{\partial q^{i_2}\cdots \partial q^{i_n}}\right|_{q = \gamma(\tau)} \,\dot\xi_1^{i_1}(\tau)\, \xi_2^{i_2}(\tau)\, \cdots\, \xi_n^{i_n}(\tau)\,d\tau \\
  \Ff_\gamma\biggl(  \begin{tikzpicture}[baseline=(X)]
    \path coordinate (A) ++(0pt,4pt) coordinate (X);
    \path (A) +(0,-8pt) node[anchor=base] {$\scriptstyle \varsigma,i$};
    \path (A) ++(20pt,0) coordinate (B) +(0,-8pt) node[anchor=base] {$\scriptstyle \tau,j$};
    \draw (A) .. controls +(0,15pt) and +(0,15pt) .. (B);
  \end{tikzpicture}  \biggr)
   = G_\gamma^{ij}(\varsigma,\tau)
\label{FeynmanRulesG}
\end{gather}
\end{defn}

We must check that the value of a marked Feynman diagram converges: the Green's function $G_\gamma(\varsigma,\tau)$ is non-smooth like $|\varsigma - \tau|$ near the diagonal, so any diagram in which both ends of the same edge are marked includes a Dirac $\delta$-function distribution.  But no vertex can have more than one marking, so such $\delta$-functions do not lead to divergence integrals.  (We have insisted that $B,C$ be smooth.  If they are allowed to be singular, divergences can occur, requiring some renormalization procedure, c.f.\ \cite{MT1994}.)  We prove the following in \cref{MainThmProof}.

\begin{thm}[Schr\"odinger's Equation]  \label{MainThm}
  Let $\gamma$ be a classical nondegenerate path, extended to a family with boundary values $(t,q_0,q_1)$ ranging over some open set $\Oo \in \RR^{2n+1}$.  Define on $\Oo$ the following $\RR[[i\hbar]]$-valued function:
  \begin{equation} \label{Vdefine}
    V_\gamma(t,q_0,q_1) = -S_\gamma(t,q_0,q_1) + i\hbar \frac12 \log\left| \det \frac{\partial^2[-S_\gamma]}{\partial q_0\partial q_1}\right| + \sum_{\substack{\text{equivalence classes }(\Gamma,M) \\ \text{of connected marked} \\ \text{Feynman diagrams}}} \frac{(i\hbar)^{\lambda(\Gamma)}\,\Ff_\gamma(\Gamma,M)}{\left|\Aut(\Gamma,M)\right|}
  \end{equation}
  Then $U_\gamma = \exp\bigl((i\hbar)^{-1}V_\gamma\bigr)$ is a formal solution to Schr\"odinger's equation: $i\hbar \frac{\partial U_\gamma}{\partial t} = \SE[U_\gamma]$ as formal power series in $\hbar$.
\end{thm}

We remark that for a connected diagram $\Gamma = (H,E,V)$, we have $\lambda(\Gamma) = |E| - |V| + 1 \leq |E| = 2|H|$, and so the sum in \cref{Vdefine} is finite at each order in $\hbar$.  In general, the series has zero radius of convergence in $\hbar$.  Note that $U_\gamma$ can be described as:
\[ U_\gamma(t,q_0,q_1) = e^{-(i\hbar)^{-1}S_\gamma} \sqrt{ \left| \det \frac{\partial[-S_\gamma]}{\partial q_0\partial q_1}\right| } \sum_{\substack{\text{equivalence classes of} \\ \text{marked Feynman} \\ \text{diagrams } (\Gamma,M)}} \frac{(i\hbar)^{-\chi(\Gamma)}\,\Ff_\gamma(\Gamma,M)}{\left|\Aut(\Gamma,M)\right|}\]

\subsection{Sums and limits of \texorpdfstring{$U_\gamma$}{U_\gamma}}

Temporarily, let $\hbar$ range over ``small'' non-zero real numbers.  Pick $N$ a positive integer and suppose that $\Oo \subseteq \RR^n$ is open, and $f_1,f_2: \Oo \to \RR[i\hbar]$ satisfy $f_1 = f_2 + O(\hbar^{N+1})$.  Let $g: \Oo \to \RR^n$ have compact support.  Then:
\begin{equation}
  \int e^{(i\hbar)^{-1}f_1(q)}\,g(q)\,dq = \int e^{(i\hbar)^{-1}f_2(q)}\,g(q)\,dq + O(\hbar^{N + n/2})
\end{equation}
For details, see \cite[Chapter 3]{EZ2007}.  Thus, if $f,g: \Oo \to \RR[[i\hbar]]$ are smooth term-by-term and $g$ has compact support, then we are justified in defining $\int \exp\bigl( (i\hbar)^{-1}\,f(q)\bigr)\,g(q)\,dq$ as a formal expression in $\hbar$, by successively approximating $f$ by polynomials in $\hbar$.  We will write $\int^\formal$ for this formal integral.

 We will not totally define what we mean by a ``formal expression in $\hbar$,'' although it is not hard to do so.  Essentially, a formal expression in $\hbar$ is an $\CC[[\sqrt{\hbar}]]$-linear combination of terms of the form $\exp\bigl((i\hbar)^{-1}f\bigr)$, where $f\in \RR[[i\hbar]]$, modulo the obvious equivalences and with the obvious arithmetic.    If $f\pmod \hbar$ has finitely many critical points within the support of $g$, and if at each critical point the Hessian of $f$ is nondegenerate, then $\int^\formal \exp\bigl((i\hbar)^{-1}f\bigr)\,g$ is a formal expression in $\hbar$ in this sense.

\begin{defn}
  Let $f_1,f_2,\dots,f_\infty$ be distributions on $\Oo \subseteq \RR^n$, valued in formal expressions in $\hbar$.  Then $\lim f_N = f_\infty$ if and only if for every $g: \Oo \to \RR$ with compact support, $\lim_{N\to \infty} \int^\formal_{\Oo} f_N(q)\,g(q)\,dq = \int^\formal_{\Oo} f_\infty(x)\,g(x)\,dx$.  This defines the \define{pointwise convergence} topology on the space of distributions.  The topology on $\RR[[i\hbar]]$ is given coefficient-by-coefficient.
\end{defn}

In \cref{IVPthmproof} we will prove:

\begin{thm}[The initial value problem]  \label{IVPthm}
  Pick $\Oo \subseteq \RR^{2n+1}$ open, and let $A$ be the set of all families of nondegenerate classical paths with boundary values varying over $\Oo$.  Let $\{g_\gamma\}_{\gamma \in A}$ be a collection of smooth functions $\Oo \to \RR[[i\hbar]]$.  Then for fixed $t,q_1$, the possibly-infinite sum
  $ \sum_{\gamma \in A}g_\gamma(t,q_0,q_1)\,\exp\bigl( (i\hbar)^{-1}V_{\gamma}(t,q_0,q_1)\bigr) $
  converges pointwise as a distribution in $q_0$ valued in formal expressions in $\hbar$.

  Moreover, suppose that $\Oo = (0,\epsilon) \times \Oo_0 \times \Oo_0$, where $\Oo_0 \subseteq \RR^n$ is open with compact closure and and $\epsilon>0$ is sufficiently small depending on $\Oo_0$.  Then for fixed $q_1$, as a distribution valued in formal expressions in $\hbar$ we have:
  \begin{equation} \label{IVPthmeqn}
    \lim_{t\to 0}  \sum_{\gamma \in A}\exp\bigl( (i\hbar)^{-1}V_{\gamma}(t,q_0,q_1)\bigr) = (2\pi  \hbar)^{n/2}e^{i n \pi / 4}\delta(q_0 - q_1)
  \end{equation}
\end{thm}

The reader should be mildly disappointed in \cref{IVPthm}: we have first set $\hbar$ to a formal parameter, and then, in \cref{IVPthmeqn}, taken a limit as $t\to 0$.  Much better would be to approximate $V_\gamma$ by a polynomial in $\hbar$, interpret $\hbar$ as a positive real parameter, take any necessary limits, and then take the asymptotics as $\hbar \to 0$.  But when the limits are performed in this order, we do not believe that $\sum_{\gamma \in A}\exp\bigl( (i\hbar)^{-1}V_{\gamma}\bigr)$ always converges.  Moreover, even if it does, some extra assumption is needed in order to take the $t\to 0$ limit for non-formal $\hbar$ in \cref{IVPthmeqn}.  For example, it is enough if the paths $\gamma \in A$ are of bounded length, although we will not prove this.  However, in as natural a system as the one-dimensional mechanics with electric potential $C(q) = q^4$, the classical paths can be arbitrarily long, and as $t\to 0$ with fixed endpoints the classical paths run away with length growing as $t^{-1/2}$.

One final remark is in order.  It is tempting to conclude from \cref{MainThm,IVPthm} that the $\hbar\to 0$ asymptotics of the full propagator $U(t,q_0,q_1)$ are be given by:
\[ U(t,q_0,q_1) \approx \sum_{\substack{\gamma\text{ classical,} \\ \gamma(0) = q_0,\,\gamma(t) = q_1}} (2\pi \hbar)^{-n/2} e^{i n \pi / 4}  \exp\bigl( (i\hbar)^{-1}\,V_\gamma(t,q_0,q_1)\bigr) \]
But a modification must be made.  The formal power series $V_\gamma$ are defined only for $\gamma$ nondegenerate, and \cref{MainThm} only applies in this case.  Thus the above formal sum solves \cref{IVPforU} only for short times in a neighborhood of the diagonal.  In general, if we know the asymptotics of $U$ near short nondegenerate paths, we can determine the full asymptotics by means of the composition law: $U(t_0+t_1, q_0, q_1) = \int U(t_0,q_0,q)\,U(t_1,q,q_1)\,dq$.  In \cite{melong} we prove the following:
\begin{thm}[Semigroup Law]
  Let $\gamma, V_\gamma$ be as in \cref{MainThm}.  Suppose that $t = t_0 + t_1$ and the restrictions $\gamma_0 = \gamma|_{[0,t_0]}$ and $\gamma_1 = \gamma|_{[t_0,t]}$ are nondegenerate, and define the corresponding formal power series $V_{\gamma_0}$ and $V_{\gamma_1}$.  Let $\eta(\gamma)$ be the Morse index of $\gamma$ with respect to the Morse function $\gamma \mapsto  \int_0^t \bigl( \frac12 \bigl| \dot\gamma\bigr|^2 + B(\gamma)\cdot \dot\gamma- C(\gamma)\bigr) d\tau$ on the space of paths with the same boundary values, and similarly define Morse indexes $\eta(\gamma_0)$ and $\eta(\gamma_1)$ (see e.g.\ \cite{Milnor1963}).  Set $U_\gamma = (-i)^{\eta(\gamma)} (2\pi \hbar)^{-n/2} e^{-i n \pi / 4}  \exp\bigl( (i\hbar)^{-1}\,V_\gamma(t,q_0,q_1)\bigr)$, and similarly define $U_{\gamma_0}$ and $U_{\gamma_1}$.  Then for a sufficiently small neighborhood $\Qq$ of $\gamma(t_0)$ we have:
  \[ U_{\gamma}(t_0 + t_1, q_0,q_1) = \int^{\formal}_\Qq U_{\gamma_0}(t_0,q_0,q)\,U_{\gamma_1}(t_1,q,q_1)\,dq \]
\end{thm}

Thus, the correct asymptotics of the full propagator $U(t,q_0,q_1)$ are essentially given by:
\[ U(t,q_0,q_1) \approx (2\pi \hbar)^{-n/2} e^{i n \pi / 4} \sum_{\substack{\gamma\text{ classical,} \\ \gamma(0) = q_0,\,\gamma(t) = q_1}} (-i)^{\eta(\gamma)}  \exp\bigl( (i\hbar)^{-1}\,V_\gamma(t,q_0,q_1)\bigr) \]
Here the sum makes sense provided that there is no degenerate classical path of duration $t$ connecting $q_0$ to $q_1$, and converges pointwise as a distribution.  This justifies the method of formal Feynman path integration.

\Section{Proofs of the Lemmas} \label{lemmatasection}

\subsection{Proof of Lemma \texorpdfstring{\ref{ChrisLemma}}{2.1.4}}  \label{ChrisLemmaProof}

Let $\Pp$ be the space of paths of duration $t$ and $\Lambda$ the space of based loops of duration $t$.  Each is an infinite-dimensional vector space; we think of $\Pp$ as an infinite-dimensional smooth manifold with tangent bundle $\T\Pp = \Pp \times \Pp$.
Recall that the \define{derivative} $df$ of a function $f: \Pp \to \RR$ is defined by $df_\gamma \cdot \xi = \left.\frac{\partial}{\partial \epsilon}\right|_{\epsilon = 0}(\gamma + \epsilon\xi)$; $f$ is \define{differentiable} if such a derivative exists for all $(\xi,\gamma) \in \T\Pp$.
 Let $\pi:\Pp \to \RR^{2n}$ be the map $\gamma \mapsto (\gamma(0),\gamma(t))$; it is a trivial bundle with fiber $\Lambda$.  As $\Lambda$ is a vector space, we can identify the fiber tangent spaces: if $\gamma\in \Pp$, then $\T_\gamma\bigl(\pi^{-1}(\pi(\gamma))\bigr) = \Lambda$.

  Consider the function $f: \Pp \to \RR$ given by:
\[ f(\gamma) = \int_0^t \left( \frac12 \bigl| \dot\gamma(\tau)\bigr|^2 + B_i\bigl(\gamma(\tau)\bigr)\,\dot\gamma^i(\tau) - C\bigl(\gamma(\tau)\bigr) \right)d\tau\]
As is well-known, a path $\gamma$ is classical if and only if $\Lambda \subseteq \ker df_\gamma$.  The \define{Hessian} of $f$ at $\gamma$ is the symmetric bilinear form:
\begin{multline*} h_\gamma:(\xi_1,\xi_2) \mapsto \int_0^t \left( \dot\xi_1^i(\tau)\,\dot\xi_2^j(\tau) + \frac{\partial B_i(q)}{\partial q^j}\,\bigl( \dot\xi_1^i(\tau)\,\xi_2^j(\tau) +\dot\xi_1^j(\tau)\,\xi_2^i(\tau) \bigr) \right. + \\ + \left.\left. \Bigl(\frac{\partial^2 B_k(q)}{\partial q^i\partial q^j}\,\dot\gamma^k(\tau) -  \frac{\partial^2 C(q)}{\partial q^i\partial q^j}\Bigr)\xi_1^i(\tau)\,\xi_2^j(\tau)  \right)\right|_{q = \gamma(\tau)}d\tau\end{multline*}
When $\xi_1,\xi_2$ are based loops, integration by parts gives $h_\gamma(\xi_1,\xi_2) = \int_0^t \xi_1(\tau) \cdot \Dd_\gamma[\xi_2](\tau)\,d\tau$.  Thus $\gamma$ is nondegenerate if and only if $\Lambda \cap \ker h_\gamma = 0$.

Let $\Cc$ be the space of classical paths.  By considering the derivatives of $df$ along paths in $\Cc$, one easily checks that $\T_\gamma\Cc \subseteq \ker h_\gamma$.  Thus if $\gamma$ is classical and nondegenerate, then $d\pi: \T_\gamma\Cc \to \T_{\pi(\gamma)}\RR^{2n}$ is an injection.  But the space of classical paths of a given duration has dimension $2n$, and so $d\pi$ is a bijection.  This completes the proof of \cref{ChrisLemma}.

\subsection{Proof of Lemma \texorpdfstring{\ref{Glemma}}{2.1.9}} \label{Gproof}

We begin with the observation that for each $a = 0,1$ and each $j = \{1,\dots,n\}$, the path $\phi_{a,j}(\tau) = \frac{\partial \gamma}{\partial q_a^j}(\tau)$ satisfies $\Dd_\gamma[\phi] = 0$; this follows by differentiating \cref{EOM}.  In other words, the $2n$ paths $\phi_{a,j}$ are \define{Jacobi fields} along $\gamma$.  The boundary values are: $\phi_{0,j}^i(0) = \delta^i_j = \phi_{1,j}^i(t)$ and $\phi_{0,j}^i(t) = \delta^i_j = \phi_{1,j}^i(0)$.  Let $\phi_0(\tau)$, $\phi_1(\tau)$ be the $n\times n$ matrices $\bigl(\phi_a(\tau)\bigr)^i_j = \phi_{a,j}^i(\tau)$.  Each solution to the initial value problem $\{\Dd_\gamma[\varphi]$, $\varphi(0) = 0\}$ is determined by $\dot\varphi(0)$.  Since $\phi_1(t) = \delta$ is full-rank, $\dot\phi_1(0)$ must be also.

Define the following function on $[0,t]$ valued in $2n \times 2n$ matrices:
\[ M(\tau) = \begin{pmatrix} \phi_0(\tau) & \phi_1(\tau) \\ \dot\phi_0(\tau) & \dot\phi_1(\tau) \end{pmatrix}\]
Then $M(0)$ is lower-triangular with full-rank blocks on the diagonal, and hence invertible.  Since the coefficient on $\dot\xi$ in $\Dd_\gamma[\xi]$ is traceless, Liouville's formula gives $\det M(\tau) = \det M(0)$, and so $M(\tau)$ is invertible for each $\tau \in [0,t]$.  Let $\psi_0,\psi_1$ be the $(n\times n)$-matrix valued functions on $[0,t]$ satisfying:
\begin{equation} \label{psidefine} \begin{pmatrix} \phi^i_{0,j}(\tau) & \phi^i_{1,j}(\tau) \\ \dot\phi^i_{0,j}(\tau) & \dot\phi^i_{1,j}(\tau) \end{pmatrix} \begin{pmatrix} \psi^j_{0,k}(\tau) \\ \psi^j_{1,k}(\tau) \end{pmatrix} = \begin{pmatrix} 0 \\ \delta^i_k \end{pmatrix} \end{equation}
Then it is straightforward to check that the Green's function for $-\Dd_\gamma$ is given by:
\[ G^i_k(\varsigma,\tau) = -\Theta(\varsigma - \tau)\, \phi^i_{0,j}(\tau)\, \psi^j_{0,k}(\varsigma) + \Theta(\tau - \varsigma) \, \phi^{1,i}_j(\tau) \, \psi^j_{1,k}(\varsigma)\]

But the differential operator $-\Dd_\gamma$ is self-adjoint for the inner product $\<\xi_1,\xi_2\> = \int_0^t \xi_1 \cdot \xi_2$ on the space of based loops.  Thus if $G^i_k(\varsigma,\tau)$ is its Green's function, then $G^i_k(\varsigma,\tau) = G^k_i(\tau,\varsigma)$.  In particular, each $\psi_a$ satisfies $\Dd_\gamma[\psi_a] = 0$, and so the $\psi_a$s are linear combinations of the $\phi_b$s.  To find the coefficients, we need only check the boundary values.  We have $\psi^j_{0,k}(0) = 0 = \psi^j_{1,k}(t)$, by evaluating the top row of \cref{psidefine} at $\tau = 0,t$.  On the other hand, the bottom row then gives:
\begin{equation} \label{Sesteqn2}
 \bigl(\psi_0(t)\bigr)^{-1} = \dot\phi_0(t) = \left.\frac{\partial^2 \gamma(\tau)}{\partial \tau\partial q_0}\right|_{\tau = t} = \frac{\partial}{\partial q_0} \bigl( \dot\gamma(t)\bigr) = \frac{\partial}{\partial q_0}\left( \frac{\partial S_\gamma}{\partial q_1} - B(q_1)\right) = \frac{\partial^2 S_\gamma}{\partial q_0\partial q_1}  \end{equation}
We have used \cref{wellknown2} to do the necessary simplifications.
Thus $\psi_{0,k}^j(\varsigma) = \phi_{1,l}^j(\varsigma)\,\Bigl( \bigl( \frac{\partial^2 S_\gamma}{\partial q_0\partial q_1}\bigr)^{-1}\Bigr)^{lk}$.  A similar argument shows that $\psi^j_{1,k}(\varsigma) = -\phi_{0,l}^j(\varsigma)\,\Bigl( \bigl( \frac{\partial^2 S_\gamma}{\partial q_1\partial q_0}\bigr)^{-1}\Bigr)^{lk}$.  This completes the proof of \cref{Glemma}.

\Section{Proof of the Main Theorem}  \label{MainThmProof}

The Schr\"odinger equation for $\exp\bigl((i\hbar)^{-1}V_\gamma\bigr)$ is equivalent to the following nonlinear differential equation for $V_\gamma$:
\begin{equation} \label{SEV}
  \frac{\partial V_\gamma}{\partial t} = \frac12 |\nabla_{q_1}V_\gamma|^2 + \frac{i\hbar}2 \,\Delta_{q_1}V_\gamma + B_j(q_1) \, \frac{\partial V_\gamma}{\partial q_1^j}  + \frac{i\hbar}2 \frac{\partial B_j(q_1)}{\partial q_1^j} + \left( \frac12 |B(q_1)|^2 + C(q_1)\right)
\end{equation}
Our strategy will be to convert \cref{SEV} into the diagrammatic language introduced earlier, whence it will follow from a few calculations of derivatives and some simple combinatorics.

\subsection{More Feynman rules}

In addition to the Feynman rules given in \crefrange{FeynmanRulesB}{FeynmanRulesG}, we introduce the following notation to our graphical calculus:\\[-2pc]
\begin{align*} \label{moreFRs}
  \Ff_\gamma\biggl(  \begin{tikzpicture}[baseline=(X)]
    \path coordinate (A) ++(0pt,4pt) coordinate (X);
    \path (A) +(0,-8pt) node[anchor=base] {$\scriptstyle i$};
    \path (A) ++(20pt,0) coordinate (B) +(0,-8pt) node[anchor=base] {$\scriptstyle j$};
    \draw[dashed] (A) .. controls +(0,15pt) and +(0,15pt) .. (B);
  \end{tikzpicture}  \biggr)
 &  = \delta^{ij} &
  \Ff_\gamma\biggl(  \tikz[baseline=(c.base)] \node[draw,inner sep=1pt,circle] (c) {$C$}; \biggr) & = C(q_1) &
  \Ff_\gamma\biggl( \tikz[baseline=(b.base)]{ \node[draw,inner sep=1pt,circle] (b) {$B$}; \draw[dashed] (b.north) -- ++(0pt,10pt) +(0pt,2pt) node[anchor=base] {$\scriptstyle i$};} \biggr) & = B_i(q_1) &
  \Ff_\gamma\biggl( \tikz[baseline=(gamma.base)]{ \node[inner sep=1pt,circle] (gamma) {$\gamma$}; \draw (gamma.south) -- ++(0pt,-10pt) +(0pt,0pt) node[anchor=north] {$\scriptstyle \tau,i$};} \biggr) & = \gamma^i(\tau)
\end{align*}
In particular, dashed lines carry only an index (no time variable).  Contraction of indices is implied by connecting dashed strands.

We will henceforth largely drop the ``$\Ff_\gamma$'' notation from equations of diagrams.  We will find it convenient to abbreviate the two kinds of vertices in marked Feynman diagrams:
\[
  \begin{tikzpicture}[baseline=(X)]
    \path node[dot] (O) {} ++(0pt,4pt) coordinate (X);
    \draw (O) -- ++(-15pt,15pt);
    \draw (O) -- ++(-10pt,15pt);
    \draw (O) -- ++(-6pt,15pt);
    \draw (O) -- ++(10pt,15pt);
    \draw (O) -- ++(15pt,15pt);
    \path (O) ++(2pt,13pt) node {$\scriptstyle \ldots$};
  \end{tikzpicture}
  =
  \begin{tikzpicture}[baseline=(X)]
    \path node[plaindot] (O) {} ++(0pt,4pt) coordinate (X);
    \draw (O) -- ++(-15pt,15pt);
    \draw (O) -- ++(-10pt,15pt);
    \draw (O) -- ++(-6pt,15pt);
    \draw (O) -- ++(10pt,15pt);
    \draw (O) -- ++(15pt,15pt);
    \path (O) ++(2pt,13pt) node {$\scriptstyle \ldots$};
  \end{tikzpicture}
  +
  \begin{tikzpicture}[baseline=(X)]
    \path node[plaindot] (O) {} ++(0pt,4pt) coordinate (X);
    \draw (O) -- node[marking]{} ++(-15pt,15pt);
    \draw (O) -- ++(-10pt,15pt);
    \draw (O) -- ++(-6pt,15pt);
    \draw (O) -- ++(10pt,15pt);
    \draw (O) -- ++(15pt,15pt);
    \path (O) ++(2pt,13pt) node {$\scriptstyle \ldots$};
  \end{tikzpicture}
  +
  \begin{tikzpicture}[baseline=(X)]
    \path node[plaindot] (O) {} ++(0pt,4pt) coordinate (X);
    \draw (O) -- ++(-15pt,15pt);
    \draw (O) -- node[marking]{} ++(-10pt,15pt);
    \draw (O) -- ++(-6pt,15pt);
    \draw (O) -- ++(10pt,15pt);
    \draw (O) -- ++(15pt,15pt);
    \path (O) ++(2pt,13pt) node {$\scriptstyle \ldots$};
  \end{tikzpicture}
  +
  \cdots
  +
  \begin{tikzpicture}[baseline=(X)]
    \path node[plaindot] (O) {} ++(0pt,4pt) coordinate (X);
    \draw (O) -- ++(-15pt,15pt);
    \draw (O) -- ++(-10pt,15pt);
    \draw (O) -- ++(-6pt,15pt);
    \draw (O) -- ++(10pt,15pt);
    \draw (O) -- node[marking]{} ++(15pt,15pt);
    \path (O) ++(2pt,13pt) node {$\scriptstyle \ldots$};
  \end{tikzpicture}
\]
Then rather than discussing marked Feynman diagrams, we will simply say ``Feynman diagram'' and mean a diagram drawn with \tikz \node[dot] {};-type vertices.  It's clear that the automorphism counts work out: if adding some markings to a diagram divides the size of the autmorphism group by some number $n$, then there were $n$ equivalent ways to add those markings to the unmarked diagram.  In particular a symmetry of a diagram with a \tikz \node[dot] {};-type vertex either acts as a symmetry on each of the expanded diagrams or permutes the possible expansions, and those that it permutes have their symmetry groups broken by the right amount.

 We will also extend the possible valence of such vertices in our notation, although not in the sums of diagrams:
\begin{gather*}
  \tikz \node[Sdot] {};  = -S_\gamma \quad\quad\quad\quad
  \begin{tikzpicture}[baseline=(X)]
    \path node[dot] (O) {} ++(0pt,4pt) coordinate (X);
    \draw (O) -- ++(0pt,15pt) +(0,3pt) node[anchor=base] {$\scriptstyle \xi$};
  \end{tikzpicture}  
   = \bigl( \dot\gamma^i(0) + B_i(q_0)\bigr)\xi^i(0) - \bigl( \dot\gamma^i(t) + B_i(q_1)\bigr)\xi^i(t)  \\
  \begin{tikzpicture}[baseline=(X)]
    \path node[dot] (O) {} ++(0pt,4pt) coordinate (X);
    \draw (O) -- ++(-10pt,15pt) +(0,3pt) node[anchor=base] {$\scriptstyle \xi$};
    \draw (O) -- ++(10pt,15pt) +(0,3pt) node[anchor=base] {$\scriptstyle \zeta$};
  \end{tikzpicture}
  =
  \begin{tikzpicture}[baseline=(X)]
    \path node[plaindot] (O) {} ++(0pt,4pt) coordinate (X);
    \draw (O) -- ++(-10pt,15pt) +(0,3pt) node[anchor=base] {$\scriptstyle \xi$};
    \draw (O) -- ++(10pt,15pt) +(0,3pt) node[anchor=base] {$\scriptstyle \zeta$};
  \end{tikzpicture}  
  +
  \begin{tikzpicture}[baseline=(X)]
    \path node[plaindot] (O) {} ++(0pt,4pt) coordinate (X);
    \draw (O) -- node[marking] {} ++(-10pt,15pt) +(0,3pt) node[anchor=base] {$\scriptstyle \xi$};
    \draw (O) -- ++(10pt,15pt) +(0,3pt) node[anchor=base] {$\scriptstyle \zeta$};
  \end{tikzpicture}  
  +
  \begin{tikzpicture}[baseline=(X)]
    \path node[plaindot] (O) {} ++(0pt,4pt) coordinate (X);
    \draw (O) -- ++(-10pt,15pt) +(0,3pt) node[anchor=base] {$\scriptstyle \xi$};
    \draw (O) -- node[marking] {} ++(10pt,15pt) +(0,3pt) node[anchor=base] {$\scriptstyle \zeta$};
  \end{tikzpicture}  
  -
  \int_0^t \dot\xi(\tau)\cdot \dot\zeta(\tau)\,d\tau
\end{gather*}
Here the test functions $\xi,\zeta$ are paths of duration $t$, and in the second line on the right-hand side the bivalent vertices have the natural meanings extending \cref{FeynmanRulesB,FeynmanRulesC} to $n=2$.  
In particular, $  \begin{tikzpicture}[baseline=(X)]
    \path node[dot] (O) {} ++(0pt,4pt) coordinate (X);
    \draw (O) -- ++(0pt,7pt) +(0,3pt) node[anchor=base] {$\scriptstyle \xi$};
  \end{tikzpicture}  
= 0$ if $\xi$ is a based loop.
By integrating by parts,  $\begin{tikzpicture}[baseline=(X)]
    \path node[dot] (O) {} ++(0pt,4pt) coordinate (X);
    \draw (O) -- ++(-5pt,7pt) +(0,3pt) node[anchor=base] {$\scriptstyle \xi$};
    \draw (O) -- ++(5pt,7pt) +(0,3pt) node[anchor=base] {$\scriptstyle \zeta$};
  \end{tikzpicture}
= \int_0^t \xi(\tau) \cdot \Dd_\gamma[\zeta](\tau)\,d\tau$ whenever $\xi$ is a based loop, where $\Dd_\gamma$ is the differential operator given in \cref{Dequation}.

We represent differentiation with respect to $t,q_0,q_1$ with dotted circles:\\[-1.5pc]
\begin{align*}
  \begin{tikzpicture}[baseline=(G.base)]
    \path node[shape=rectangle,draw,inner sep=2pt] (G) {$\Gamma$} ++(12pt,0) coordinate (H);
    \draw (G.west) ++(0,1pt) .. controls +(-10pt,0pt) and +(5pt,-5pt) .. ++(-25pt,20pt);
    \draw (G.east) ++(0,1pt) .. controls +(10pt,0pt) and +(-5pt,-5pt) .. ++(25pt,20pt);
    \draw (G.east) ++(0,-1pt) .. controls +(10pt,0pt) and +(-5pt,5pt) .. ++(25pt,-20pt);
    \node [draw,dotted,thick, circle through=(H)] (C) at (G) {};
    \node[anchor=south west] at (C.north) {$\scriptstyle t$};
  \end{tikzpicture}
  & = \frac{\partial}{\partial t} \bigl[ \Ff(\Gamma)\bigr] &
  \begin{tikzpicture}[baseline=(G.base)]
    \path node[shape=rectangle,draw,inner sep=2pt] (G) {$\Gamma$} ++(12pt,0) coordinate (H);
    \draw (G.west) ++(0,1pt) .. controls +(-10pt,0pt) and +(5pt,-5pt) .. ++(-25pt,20pt);
    \draw (G.east) ++(0,1pt) .. controls +(10pt,0pt) and +(-5pt,-5pt) .. ++(25pt,20pt);
    \draw (G.east) ++(0,-1pt) .. controls +(10pt,0pt) and +(-5pt,5pt) .. ++(25pt,-20pt);
    \node [draw,dotted,thick, circle through=(H)] (C) at (G) {};
    \node[anchor=south west] at (C.north) {$\scriptstyle q_a$};
    \draw [dashed] (C.north) -- ++(0,15pt) +(0,3pt) node[anchor=base] {$\scriptstyle i$};
  \end{tikzpicture}
  & = \frac{\partial}{\partial q_a^i} \bigl[ \Ff(\Gamma)\bigr] 
\end{align*}
Then the product rule can be written graphically as:
\begin{align*}
  \begin{tikzpicture}[baseline=(D.base)]
    \node (D) {$\scriptstyle \cdots$};
    \path (D) ++(-3pt,25pt) node[shape=rectangle,draw,inner sep=2pt] (G1) {$\Gamma_1$};
    \path (D) ++(-3pt,-25pt) node[shape=rectangle,draw,inner sep=2pt] (G2) {$\Gamma_2$};
    \draw (G1) .. controls +(-4pt,-15pt) and +(-4pt,15pt) .. (G2);
    \draw (G1) .. controls +(-10pt,-15pt) and +(-10pt,15pt) .. (G2);
    \draw (G1) .. controls +(10pt,-15pt) and +(10pt,15pt) .. (G2);
    \draw[dotted, thick] (D) ++(-3pt,0pt) ellipse (20pt and 40pt);
  \end{tikzpicture}
  =
  \begin{tikzpicture}[baseline=(D.base)]
    \node (D) {$\scriptstyle \cdots$};
    \path (D) ++(-3pt,25pt) node[shape=rectangle,draw,inner sep=2pt] (G1) {$\Gamma_1$};
    \path (D) ++(-3pt,-25pt) node[shape=rectangle,draw,inner sep=2pt] (G2) {$\Gamma_2$};
    \draw (G1) .. controls +(-4pt,-15pt) and +(-4pt,15pt) .. (G2);
    \draw (G1) .. controls +(-10pt,-15pt) and +(-10pt,15pt) .. (G2);
    \draw (G1) .. controls +(10pt,-15pt) and +(10pt,15pt) .. (G2);
    \draw[dotted, thick] (G1) circle (15pt);
  \end{tikzpicture}
  +
  \begin{tikzpicture}[baseline=(D.base)]
    \node (D) {$\scriptstyle \cdots$};
    \path (D) ++(-3pt,25pt) node[shape=rectangle,draw,inner sep=2pt] (G1) {$\Gamma_1$};
    \path (D) ++(-3pt,-25pt) node[shape=rectangle,draw,inner sep=2pt] (G2) {$\Gamma_2$};
    \draw (G1) .. controls +(-4pt,-15pt) and +(-4pt,15pt) .. (G2);
    \draw (G1) .. controls +(-10pt,-15pt) and +(-10pt,15pt) .. (G2);
    \draw (G1) .. controls +(10pt,-15pt) and +(10pt,15pt) .. (G2);
    \draw[dotted, thick] (G2) circle (15pt);
  \end{tikzpicture}  
\end{align*}
Suppose then that $\Gamma_1$ is a subdiagram of $\Gamma$ whose images $\Gamma_1,\dots,\Gamma_n$ under the group of automorphisms of $\Gamma$ do not intersect, so that $\Gamma = \bar\Gamma \cup \Gamma_1\cup \dots \cup \Gamma_n$ (we do allow automorphism of $\Gamma$ to induce nontrivial automorphisms of $\Gamma_1$).  Then: 
\begin{multline*}
  \tikz[baseline=(G.base)]    \node (G) [shape=ellipse,draw,dotted,thick,inner sep=3pt] {$\Gamma$};
   = 
  \tikz[baseline=(G.base)]    \node (G) [shape=ellipse,draw,dotted,thick,inner sep=3pt] {$\bar \Gamma\,\Gamma_1\,\Gamma_2\cdots\Gamma_n$};
  = 
  \tikz[baseline=(G.base)]    \node (G) [shape=ellipse,draw,dotted,thick,inner sep=2pt] {$\bar \Gamma$}; \,\Gamma_1\,\Gamma_2\cdots\Gamma_n
  +
  \bar \Gamma\, \tikz[baseline=(G.base)]    \node (G) [shape=ellipse,draw,dotted,thick,inner sep=2pt] {$\Gamma_1$};\,\Gamma_2\cdots\Gamma_n
  + \cdots +
  \bar \Gamma\,\Gamma_1\,\Gamma_2\,\cdots\,\tikz[baseline=(G.base)]    \node (G) [shape=ellipse,draw,dotted,thick,inner sep=2pt] {$\Gamma_n$};
  = \\ = 
  \tikz[baseline=(G.base)]    \node (G) [shape=ellipse,draw,dotted,thick,inner sep=2pt] {$\bar \Gamma$}; \,\Gamma_1\,\Gamma_2\cdots\Gamma_n
  +
  n\,\bar \Gamma\, \tikz[baseline=(G.base)]    \node (G) [shape=ellipse,draw,dotted,thick,inner sep=2pt] {$\Gamma_1$};\,\Gamma_2\cdots\Gamma_n
\end{multline*}
It is an elementary counting lemma that $
  \Aut\left( \bar \Gamma\, \tikz[baseline=(G.base)]    \node (G) [shape=ellipse,draw,dotted,thick,inner sep=2pt] {$\Gamma_1$};\,\Gamma_2\cdots\Gamma_n\right) = \frac1n \Aut\left( \bar \Gamma\,\Gamma_1\,\Gamma_2\cdots\Gamma_n\right)
$.
  From this observation, we derive the fundamental result that for $z = q_1^i$ or $t$: 
\begin{equation} 
  \label{productrule}
  \frac{\partial}{\partial z} \sum_{\Gamma \textrm{ connected}} \frac{(i\hbar)^{\lambda(\Gamma)}\Ff_\gamma(\Gamma)}{\left| \Aut \Gamma\right|} = \sum_{\substack{\Gamma \textrm{ connected with} \\ \textrm{one \, \begin{tikzpicture}[baseline=(O.base)]
    \node[dot] (O) {};
    \draw (O) -- ++(-5pt,8pt);
    \draw (O) -- ++(-2pt,9pt);
    \node at (1pt,9pt) {.};
    \node at (3pt,9pt) {.};
    \node at (5pt,9pt) {.};
    \draw (O) -- ++(6pt,7pt);
    \draw[dotted,thick] (O) circle (5pt);
  \end{tikzpicture}
   or \,
  \begin{tikzpicture}[baseline=(O.base)]
    \path coordinate (O) ++(0,-5pt) coordinate (A) ++(10pt,0) coordinate (B);
    \draw (A) .. controls +(0,10pt) and +(0,10pt) .. coordinate (C) (B);
    \draw[dotted,thick] (C) circle (4pt);    
  \end{tikzpicture}}}} \frac{(i\hbar)^{\lambda(\Gamma)}\Ff_\gamma(\Gamma)}{\left|\Aut \Gamma\right|}
\end{equation}
The sum on the right-hand side ranges over connected components with precisely one differentiated basic subgraph --- either a single differentiated vertex (of valence three or more) or a single differentiated edge.

In this notation, \cref{wellknown1} reads:
\begin{gather} \label{HJ}
  \tikz[baseline=(S.south)]{\node[Sdot] (S) {}; \node (C) [draw,circle,dotted,thick,inner sep=2pt, fit=(S)] {}; \path (C.north west) node[anchor=south] {$\scriptstyle t$};}
  \, = \,
  \frac12 \, 
   \begin{tikzpicture}[baseline=(S1.south)]
    \path node[Sdot] (S1) {} +(0,0pt) coordinate (X);
    \path (S1) ++(40pt,0pt) node[Sdot] (S2) {};
    \node[draw,dotted,thick,inner sep=2pt,circle,fit =(S1)] (C1) {};        
    \node[draw,dotted,thick,inner sep=2pt,circle,fit =(S2)] (C2) {};        
    \draw[dashed] (C1) .. controls +(10pt,20pt) and +(-10pt,20pt) .. (C2);
    \path (C1.north) node[anchor=south] {$\scriptstyle q_1$};
    \path (C2.north) node[anchor=south] {$\scriptstyle q_1$};
  \end{tikzpicture}
  \, + \,
   \begin{tikzpicture}[baseline=(S1.south)]
    \path node[Sdot] (S1) {} +(0,7pt) coordinate (X);
    \path (S1) ++(40pt,0pt) node[inner sep=1pt,circle,draw] (b) {$B$};
    \node[draw,dotted,thick,inner sep=2pt,circle,fit =(S1)] (C1) {};        
    \draw[dashed] (C1) .. controls +(10pt,20pt) and +(-10pt,20pt) .. (b);
    \path (C1.north) node[anchor=south] {$\scriptstyle q_1$};
  \end{tikzpicture}
  \, + \,
  \frac12 \, 
   \begin{tikzpicture}[baseline=(S.south)]
    \path node[rectangle,inner sep=2pt] (S) {} node[inner sep=1pt,circle,draw] (b1) {$B$} +(0,7pt) coordinate (X);
    \path (b1) ++(40pt,0pt) node[inner sep=1pt,circle,draw] (b) {$B$};
    \draw[dashed] (b1) .. controls +(10pt,20pt) and +(-10pt,20pt) .. (b);
  \end{tikzpicture}  
  \, + \,
  \tikz[baseline=(S.south)] \path node[rectangle,inner sep=2pt] (S) {} node[draw,inner sep=1pt,ellipse] (c) {$C$};  
\end{gather}
Moreover, it's clear that $\frac{\partial \gamma(\tau)}{\partial t} = - \frac{\partial \gamma(\tau)}{\partial q_1^i}\,\dot\gamma^i(t)$.  Given \cref{wellknown2}, we have:
\begin{equation} \label{dgammadt}
  \begin{tikzpicture}[baseline=(gamma.base)]
    \path node[inner sep=1pt] (gamma) {$\gamma$};
    \draw (gamma) -- ++(0,-15pt);
    \path (gamma) node (C) [draw,circle,dotted,thick,inner sep=5pt] {};
    \path (C.north) ++(1pt,3pt) node[anchor= east] {$\scriptstyle t$};
  \end{tikzpicture}
  \,  =\, 
  \begin{tikzpicture}[baseline=(gamma.base)]
    \path node[inner sep=0pt] (gamma) {$\gamma$};
    \path (gamma) ++(40pt,0pt) node[Sdot] (S) {};
    \draw (gamma) -- ++(-10pt,-15pt);
    \node[draw,dotted,thick,inner sep=2pt,circle,fit =(gamma)] (C1) {};        
    \node[draw,dotted,thick,inner sep=2pt,circle,fit =(S)] (C2) {};        
    \draw[dashed] (C1) .. controls +(10pt,20pt) and +(-10pt,20pt) .. (C2);
    \path (C1.north) node[anchor=south] {$\scriptstyle q_1$};
    \path (C2.north) node[anchor=south] {$\scriptstyle q_1$};
  \end{tikzpicture}
  \, + \,
    \begin{tikzpicture}[baseline=(gamma.base)]
    \path node[inner sep=0pt] (gamma) {$\gamma$};
    \path (gamma) ++(40pt,0pt) node[inner sep=1pt,circle,draw] (b) {$B$};
    \draw (gamma) -- ++(-10pt,-15pt);
    \node[draw,dotted,thick,inner sep=2pt,circle,fit =(gamma)] (C1) {};        
    \draw[dashed] (C1) .. controls +(10pt,20pt) and +(-10pt,20pt) .. (b);
    \path (C1.north) node[anchor=south] {$\scriptstyle q_1$};
  \end{tikzpicture}
\end{equation}

Finally, we introduce the Feynman rule $_0\tikz[baseline=(S.base)] \node[draw,ellipse,inner sep=1pt] (S) {$(\partial^2[-S])^{-1}$};{}_1$ for the inverse matrix $\bigl( \frac{\partial^2[-S]}{\partial q_0\partial q_1}\bigr)^{-1}$:
\begin{equation*}
  \begin{tikzpicture}[baseline=(c)]
    \path coordinate (b) ++(0pt,40pt) node[inner sep=1pt,ellipse,draw] (S1) {$(\partial^2[-S])^{-1}$} ++(0,20pt) coordinate (t) ++(0,-30pt) coordinate (c);
    \draw[dashed] (S1.south west) -- (S1.south west |- b);
    \path (S1.south east) ++(20pt,-20pt) node[Sdot] (S) {};
    \node (C1) [draw,circle,dotted,thick,inner sep=2pt, fit=(S)] {};    
    \node (C2) [draw,circle,dotted,thick,inner sep=3pt, fit=(S)] {};    
    \draw[dashed] (S1.south east) -- (C2.north west);
    \draw[dashed] (C2.north east) -- (C2.north east |- t);
     \path (S1.south east) ++(1pt,-2pt) node[anchor= west] {$\scriptstyle 1$};
    \path (S1.south west) ++(0pt,-3pt) node[anchor= east] {$\scriptstyle 0$};
    \path (C2.north west) node[anchor=east] {$\scriptstyle q_1$};
    \path (C2.north east) node[anchor=west] {$\scriptstyle q_0$};
  \end{tikzpicture}
  = \quad
  \begin{tikzpicture}[baseline=(c)]
    \path coordinate (b) ++(0pt,60pt) coordinate (t) ++(0,-30pt) coordinate (c);
    \draw[dashed] (b) -- (t);
  \end{tikzpicture}
  \quad =
  \begin{tikzpicture}[baseline=(c)]
    \path coordinate (b) ++(0pt,40pt) node[inner sep=1pt,ellipse,draw] (S1) {$(\partial^2[-S])^{-1}$} ++(0,20pt) coordinate (t) ++(0,-30pt) coordinate (c);
    \draw[dashed] (S1.south east) -- (S1.south east |- b);
    \path (S1.south west) ++(-20pt,-20pt) node[Sdot] (S) {};
    \node (C1) [draw,circle,dotted,thick,inner sep=2pt, fit=(S)] {};    
    \node (C2) [draw,circle,dotted,thick,inner sep=3pt, fit=(S)] {};    
    \draw[dashed] (S1.south west) -- (C2.north east);
    \draw[dashed] (C2.north west) -- (C2.north west |- t);
     \path (S1.south east) ++(0pt,-3pt) node[anchor= west] {$\scriptstyle 1$};
    \path (S1.south west) ++(-1pt,-2pt) node[anchor= east] {$\scriptstyle 0$};
    \path (C2.north west) node[anchor=east] {$\scriptstyle q_1$};
    \path (C2.north east) node[anchor=west] {$\scriptstyle q_0$};
  \end{tikzpicture}
\end{equation*}
For example, in this notation \cref{Gdefine} reads:
\begin{equation} \label{greeneqn2}
  \begin{tikzpicture}[baseline=(X)]
    \path coordinate (A) ++(0pt,2pt) coordinate (X);
    \path (A) +(0,-8pt) node[anchor=base] {$\scriptstyle \varsigma$};
    \path (A) ++(20pt,0) coordinate (B) +(0,-8pt) node[anchor=base] {$\scriptstyle \tau$};
    \draw (A) .. controls +(0,15pt) and +(0,15pt) .. (B);
  \end{tikzpicture}  
  =
  \begin{tikzpicture}[baseline=(X)]
    \path node[inner sep=1pt,ellipse,draw] (S) {$(\partial^2[-S])^{-1}$} +(0,-20pt) coordinate (X);
    \path (S.south west) ++(0pt,-20pt) node (gamma1) {$\gamma$};
    \path (S.south east) ++(0pt,-20pt) node (gamma2) {$\gamma$};
    \path (S.south east) ++(0pt,-3pt) node[anchor= west] {$\scriptstyle 1$};
    \path (S.south west) ++(0pt,-3pt) node[anchor= east] {$\scriptstyle 0$};
    \draw (gamma1) -- ++(0,-15pt) +(0,-6pt) node[anchor=base] {$\scriptstyle \varsigma$};
    \path (gamma1) node (C1) [draw,circle,dotted,thick,inner sep=5pt] {};
    \path (C1.north) ++(1pt,3pt) node[anchor= west] {$\scriptstyle q_0$};
    \draw (gamma2) -- ++(0,-15pt) +(0,-6pt) node[anchor=base] {$\scriptstyle \tau$};
    \path (gamma2) node (C2) [draw,circle,dotted,thick,inner sep=5pt] {};
    \path (C2.north) ++(1pt,3pt) node[anchor= east] {$\scriptstyle q_1$};
    \draw[dashed] (C1.north) -- (S.south west);
    \draw[dashed] (C2.north) -- (S.south east);
  \end{tikzpicture}
  \;
  \Theta(\varsigma - \tau)
  \;
  +
  \;
  \begin{tikzpicture}[baseline=(X)]
    \path node[inner sep=1pt,ellipse,draw] (S) {$(\partial^2[-S])^{-1}$} +(0,-20pt) coordinate (X);
    \path (S.south west) ++(0pt,-20pt) node (gamma1) {$\gamma$};
    \path (S.south east) ++(0pt,-20pt) node (gamma2) {$\gamma$};
    \path (S.south east) ++(0pt,-3pt) node[anchor= west] {$\scriptstyle 0$};
    \path (S.south west) ++(0pt,-3pt) node[anchor= east] {$\scriptstyle 1$};
    \draw (gamma1) -- ++(0,-15pt) +(0,-6pt) node[anchor=base] {$\scriptstyle \varsigma$};
    \path (gamma1) node (C1) [draw,circle,dotted,thick,inner sep=5pt] {};
    \path (C1.north) ++(1pt,3pt) node[anchor= west] {$\scriptstyle q_1$};
    \draw (gamma2) -- ++(0,-15pt) +(0,-6pt) node[anchor=base] {$\scriptstyle \tau$};
    \path (gamma2) node (C2) [draw,circle,dotted,thick,inner sep=5pt] {};
    \path (C2.north) ++(1pt,3pt) node[anchor= east] {$\scriptstyle q_0$};
    \draw[dashed] (C1.north) -- (S.south west);
    \draw[dashed] (C2.north) -- (S.south east);
  \end{tikzpicture}
  \;
  \Theta(\tau - \varsigma)
\end{equation}

\subsection{Derivatives of the vertices}

In this section we describe the derivatives of the vertices.  We begin with the derivatives with respect to $q_a$.
\Cref{wellknown2} implies:
\begin{equation*}
  \begin{tikzpicture}[baseline=(X)]
    \path node[Sdot] (S) {};
    \path (S) ++(0pt,20pt) coordinate (X) ++(0pt,20pt) coordinate (t);
    \node (C) [draw,circle,dotted,thick,inner sep=2pt, fit=(S)] {};
    \path (C.north) ++(1pt,3pt) node[anchor= east] {$\scriptstyle q_a$};
    \draw[dashed] (C.north) -- (t)  +(0,3pt) node[anchor=base] {$\scriptstyle i$};
  \end{tikzpicture}
  =
  \begin{tikzpicture}[baseline=(X)]
    \path node[Sdot] (S) {} ++(0,20pt) node[inner sep=1pt] (gamma) {$\gamma$};
    \path (S) ++(0pt,20pt) coordinate (X) ++(0pt,20pt) coordinate (t);
    \draw (S) -- (gamma);
    \node (C) [draw,circle,dotted,thick,inner sep=1pt, fit=(gamma)] {};
    \draw[dashed] (C) -- (t) +(0,3pt) node[anchor=base] {$\scriptstyle i$};
    \path (C.north) ++(1pt,3pt) node[anchor= east] {$\scriptstyle q_a$};
  \end{tikzpicture}
\end{equation*}
Integration by parts and the observations in the first paragraph of \cref{Gproof} give:
\[ 
  \begin{tikzpicture}[baseline=(X)]
    \path node[Sdot] (S) {} ++(-15pt,20pt) node[inner sep=0pt] (gamma) {$\gamma$};
    \path (S) ++(0pt,20pt) coordinate (X) ++(-15pt,20pt) coordinate (t);
    \draw (S) -- (gamma);
    \node (C) [draw,circle,dotted,thick,inner sep=2pt, fit=(gamma)] {};
    \draw[dashed] (C) -- (t) +(0,3pt) node[anchor=base] {$\scriptstyle i$};
    \path (C.north) ++(1pt,3pt) node[anchor= east] {$\scriptstyle q_a$};
    \draw (S) -- ++(15pt,40pt) +(0,3pt) node[anchor=base] {$\scriptstyle \xi$};
  \end{tikzpicture}
  =
  \int_0^t \Dd_\gamma\left[ \frac{\partial\gamma(\tau)}{\partial q_a}\right] \cdot \xi(\tau)\,d\tau + \left.\left( - \frac{\partial\dot\gamma^j(\tau)}{\partial q_a^i} \xi^j(\tau) - \frac{\partial B_k}{\partial q^j} \frac{\partial\gamma^k(\tau)}{\partial q_a^i} \xi^j(\tau) \right) \right|_{\tau=0}^t = 
  \begin{tikzpicture}[baseline=(X)]
    \path node[Sdot] (S) {};
    \path (S) ++(0pt,20pt) coordinate (X) ++(-15pt,20pt) coordinate (t);
    \node (C) [draw,circle,dotted,thick,inner sep=2pt, fit=(S)] {};
    \draw[dashed] (C) -- (t) +(0,3pt) node[anchor=base] {$\scriptstyle i$};
    \path (C.north) ++(1pt,3pt) node[anchor= east] {$\scriptstyle q_a$};
    \draw (S) -- ++(15pt,40pt) +(0,3pt) node[anchor=base] {$\scriptstyle \xi$};
  \end{tikzpicture}
\]

In valence $n\geq 2$, the derivatives are particularly easy.  Differentiating under the integral sign and using the chain rule gives:\\[-1.5pc]
\begin{equation*}
  \begin{tikzpicture}[baseline=(X)]
    \path node[plaindot] (O) {} ++(0pt,4pt) coordinate (X);
    \draw (O) -- node[marking] {} ++(-15pt,15pt) +(0,3pt) node[anchor=base] {$\scriptstyle \xi_1$};
    \draw (O) -- ++(-6pt,15pt) +(0,3pt) node[anchor=base] {$\scriptstyle \xi_2$};
    \draw (O) -- ++(15pt,15pt) +(0,3pt) node[anchor=base] {$\scriptstyle \xi_n$};
    \path (O) ++(4pt,13pt) node {$\scriptstyle \ldots$};
    \path (O) node (C) [draw,circle,dotted,thick,inner sep=5pt] {};
    \draw[dashed] (C) .. controls +(-10pt,5pt) and +(3pt,-10pt) .. ++(-30pt,25pt) +(0,3pt) node[anchor=base] {$\scriptstyle j$};
    \path (C.west) node[anchor=east] {$\scriptstyle q_a$};
  \end{tikzpicture}
  = \int_{\tau = 0}^t \left.\frac{\partial}{\partial q^k} \frac{\partial^{n-1} [-B_{i_1}]}{\partial q^{i_2}\cdots \partial q^{i_n}}\right|_{q = \gamma(\tau)}\, \frac{\partial\gamma^k(\tau)}{\partial q_a^j} \,\dot\xi_1^{i_1}(\tau)\, \xi_2^{i_2}(\tau)\, \cdots\, \xi_n^{i_n}(\tau)\,d\tau = 
  \begin{tikzpicture}[baseline=(X)]
    \path node[plaindot] (O) {} ++(0pt,4pt) coordinate (X);
    \draw (O) -- node[marking] {} ++(-15pt,15pt) +(0,3pt) node[anchor=base] {$\scriptstyle \xi_1$};
    \draw (O) -- ++(-6pt,15pt) +(0,3pt) node[anchor=base] {$\scriptstyle \xi_2$};
    \draw (O) -- ++(15pt,15pt) +(0,3pt) node[anchor=base] {$\scriptstyle \xi_n$};
    \path (O) ++(4pt,13pt) node {$\scriptstyle \ldots$};
    \path (O) ++(-30pt,10pt) node[inner sep=1pt] (gamma) {$\gamma$};
    \draw (O) -- (gamma);
    \path (gamma) node (C) [draw,circle,dotted,thick,inner sep=5pt] {};
    \draw[dashed] (C) -- ++(0,18pt) +(0,3pt) node[anchor=base] {$\scriptstyle j$};
    \path (C.north) ++(1pt,3pt) node[anchor= east] {$\scriptstyle q_a$};
  \end{tikzpicture}
\end{equation*}
A similar argument gives:\\[-1.5pc]
\begin{equation*}
  \begin{tikzpicture}[baseline=(X)]
    \path node[plaindot] (O) {} ++(0pt,4pt) coordinate (X);
    \draw (O) -- ++(-15pt,15pt) +(0,3pt) node[anchor=base] {$\scriptstyle \xi_1$};
    \draw (O) -- ++(-6pt,15pt) +(0,3pt) node[anchor=base] {$\scriptstyle \xi_2$};
    \draw (O) -- ++(15pt,15pt) +(0,3pt) node[anchor=base] {$\scriptstyle \xi_n$};
    \path (O) ++(4pt,13pt) node {$\scriptstyle \ldots$};
    \path (O) node (C) [draw,circle,dotted,thick,inner sep=5pt] {};
    \draw[dashed] (C) .. controls +(-10pt,5pt) and +(3pt,-10pt) .. ++(-30pt,25pt) +(0,3pt) node[anchor=base] {$\scriptstyle j$};
    \path (C.west) node[anchor=east] {$\scriptstyle q_a$};
  \end{tikzpicture}
  =
  \begin{tikzpicture}[baseline=(X)]
    \path node[plaindot] (O) {} ++(0pt,4pt) coordinate (X);
    \draw (O) -- ++(-15pt,15pt) +(0,3pt) node[anchor=base] {$\scriptstyle \xi_1$};
    \draw (O) -- ++(-6pt,15pt) +(0,3pt) node[anchor=base] {$\scriptstyle \xi_2$};
    \draw (O) -- ++(15pt,15pt) +(0,3pt) node[anchor=base] {$\scriptstyle \xi_n$};
    \path (O) ++(4pt,13pt) node {$\scriptstyle \ldots$};
    \path (O) ++(-30pt,10pt) node[inner sep=1pt] (gamma) {$\gamma$};
    \draw (O) -- (gamma);
    \path (gamma) node (C) [draw,circle,dotted,thick,inner sep=5pt] {};
    \draw[dashed] (C) -- ++(0,18pt) +(0,3pt) node[anchor=base] {$\scriptstyle j$};
    \path (C.north) ++(1pt,3pt) node[anchor= east] {$\scriptstyle q_a$};
  \end{tikzpicture}
  +
  \begin{tikzpicture}[baseline=(X)]
    \path node[plaindot] (O) {} ++(0pt,4pt) coordinate (X);
    \draw (O) -- ++(-15pt,15pt) +(0,3pt) node[anchor=base] {$\scriptstyle \xi_1$};
    \draw (O) -- ++(-6pt,15pt) +(0,3pt) node[anchor=base] {$\scriptstyle \xi_2$};
    \draw (O) -- ++(15pt,15pt) +(0,3pt) node[anchor=base] {$\scriptstyle \xi_n$};
    \path (O) ++(4pt,13pt) node {$\scriptstyle \ldots$};
    \path (O) ++(-30pt,10pt) node[inner sep=1pt] (gamma) {$\gamma$};
    \draw (O) -- node[marking] {}  (gamma);
    \path (gamma) node (C) [draw,circle,dotted,thick,inner sep=5pt] {};
    \draw[dashed] (C) -- ++(0,18pt) +(0,3pt) node[anchor=base] {$\scriptstyle j$};
    \path (C.north) ++(1pt,3pt) node[anchor= east] {$\scriptstyle q_a$};
  \end{tikzpicture}
\end{equation*}
Summing and noting also that the non-diagrammatic part of the bivalent vertex does not depend on the $q_a$s, we have, for all $n$:\\[-2pc]
\begin{equation} \label{dAdQ}
  \begin{tikzpicture}[baseline=(X)]
    \path node[dot] (O) {} ++(0pt,4pt) coordinate (X);
    \draw (O) -- ++(-15pt,15pt) +(0,3pt) node[anchor=base] {$\scriptstyle \xi_1$};
    \draw (O) -- ++(-6pt,15pt) +(0,3pt) node[anchor=base] {$\scriptstyle \xi_2$};
    \draw (O) -- ++(15pt,15pt) +(0,3pt) node[anchor=base] {$\scriptstyle \xi_n$};
    \path (O) ++(4pt,13pt) node {$\scriptstyle \ldots$};
    \path (O) node (C) [draw,circle,dotted,thick,inner sep=5pt] {};
    \draw[dashed] (C) .. controls +(-10pt,5pt) and +(3pt,-10pt) .. ++(-30pt,25pt) +(0,3pt) node[anchor=base] {$\scriptstyle j$};
    \path (C.west) node[anchor=east] {$\scriptstyle q_a$};
  \end{tikzpicture}
  =
  \begin{tikzpicture}[baseline=(X)]
    \path node[dot] (O) {} ++(0pt,4pt) coordinate (X);
    \draw (O) -- ++(-15pt,15pt) +(0,3pt) node[anchor=base] {$\scriptstyle \xi_1$};
    \draw (O) -- ++(-6pt,15pt) +(0,3pt) node[anchor=base] {$\scriptstyle \xi_2$};
    \draw (O) -- ++(15pt,15pt) +(0,3pt) node[anchor=base] {$\scriptstyle \xi_n$};
    \path (O) ++(4pt,13pt) node {$\scriptstyle \ldots$};
    \path (O) ++(-30pt,10pt) node[inner sep=1pt] (gamma) {$\gamma$};
    \draw (O) -- (gamma);
    \path (gamma) node (C) [draw,circle,dotted,thick,inner sep=5pt] {};
    \draw[dashed] (C) -- ++(0,18pt) +(0,3pt) node[anchor=base] {$\scriptstyle j$};
    \path (C.north) ++(1pt,3pt) node[anchor= east] {$\scriptstyle q_a$};
  \end{tikzpicture}
\end{equation}

The derivatives with respect to $t$ are harder because of the boundary terms.  At valence $n=0$,  \cref{HJ} already describes \!\tikz[baseline=(S.south)]{\node[Sdot] (S) {}; \node (C) [draw,circle,dotted,thick,inner sep=2pt, fit=(S)] {}; \path (C.north west) node[anchor=east] {$\scriptstyle t$};}.  We will not describe the other $t$-derivatives in full generality, as we will only need them for valence $n\geq 3$ when all incoming edges are based loops.  In this case, when $n\geq 3$ at least one incoming edge is not marked, and hence the boundary terms vanish.  We have:
\begin{equation}

  , \;\; n \geq 3,\text{ all $\xi_i$s are based loops}
\end{equation}

We conclude this section by addressing the second derivative of a vertex with respect to the $q_a$.  By the product rule:\\[-2pc]
\begin{align*}
  %
 
  = 0 \text{ if $\xi$ is a based loop}.
\end{equation*}
The Green's function $G_\gamma$, which we denote by an edge, is a based loop in each variable, so:\\[-.5pc]
\begin{equation*}  0  = 
  %
 \label{contractedeqn}
\end{equation*}
We have observed already that when contracted with based loops, the bivalent vertex acts as $\Dd_\gamma$.  But $\Dd_\gamma[G_\gamma]^i_j(\varsigma,\tau) = -\delta^i_j\delta(\varsigma-\tau)$ by definition.  In short-hand: \\[-1pc]
\begin{equation}
  %
     
  \text{, provided $\xi$ is a based loop.}
\label{greendefngraphical} \end{equation}
And $\frac{\partial^2\gamma(\tau)}{\partial q_a\partial q_b}$ vanishes at both endpoints $\tau = 0,t$.  Therefore:\\[-1pc]
\begin{equation}
  \frac{\partial^2 \gamma^i(\tau)}{\partial q_a^{j}\partial q_b^{k}} = 
  %
   
\label{dphithree} \end{equation}

Thus in general to take the second derivative of a vertex one either adds two $\frac{\partial \gamma}{\partial q_a}$s or an edge connecting to a trivalent vertex with two $\frac{\partial\gamma}{\partial q_a}$s.  The lowest-valence example is:\\[-.5pc]
\begin{equation}\label{dSdQQ}
  %
    
  \, + \, 0
\end{equation}
The second summand vanishes because $G_\gamma$ is a based loop in each variable.  As we will need it later, we record one further derivative:
\begin{multline}
  %
 
  \, + 0 \label{dSdQQQ}
\end{multline}

\subsection{Derivatives of the Green's function and the determinant}

In this section, we evaluate the derivatives of the Green's function $G_\gamma$ given by \cref{Gdefine} with respect to $q_a,t$.  We record also the derivatives of $\log \left| \det \frac{\partial^2[-S_\gamma]}{\partial q_0\partial q_1}\right|$.

It is easy to evaluate $\frac{\partial G_\gamma}{\partial q_a}$.  Let $\xi$ be a based loop, and consider the derivative of \cref{greendefngraphical}:\\[-1.5pc]
\begin{align} 0 =
  %
  \label{derGreen1} \end{align}
We contract with $G_\gamma$, which is a based loop in each variable, and use \cref{dAdQ,greendefngraphical}:\\[-.5pc]
\begin{equation}
  %
      \label{dGdQ}
\end{equation}
The first equality in \cref{dGdQ} requires that $\frac{\partial G_\gamma}{\partial q_a}$ is a based loop, which follows from differentiating $G_\gamma(\varsigma,0) = 0 = G_\gamma(\varsigma,t)$ with respect to $q_a$.

Evaluating $\frac{\partial G}{\partial t}$ is not so easy, because it is not a based loop: from \cref{dGdT} it will follow that $\frac{\partial G^{ij}}{\partial t}(t,t) = \delta^{ij}$.   Instead, we will differentiate \cref{greeneqn2} directly.   To begin, we evaluate $\frac{\partial^2\gamma}{\partial t\partial q_a}$ by differentiating \cref{dgammadt} with respect to $q_a$, using \cref{dphithree,dSdQQ}:
\begin{align}
  \begin{tikzpicture}[baseline=(gamma.base)]
    \path node[inner sep=1pt] (gamma) {$\gamma$};
    \draw (gamma) -- ++(0,-15pt);
    \path (gamma) node [draw,circle,dotted,thick,inner sep=5pt] {};
    \path (gamma) node (C) [draw,circle,dotted,thick,inner sep=6pt] {};
    \path (C.west) ++(1pt,0pt) node[anchor= east] {$\scriptstyle t$};
    \path (C.north) ++(-1pt,3pt) node[anchor= west] {$\scriptstyle q_a$};
    \draw[dashed] (C.north) -- ++(0,20pt);
  \end{tikzpicture}
  \, & =\, 
  \begin{tikzpicture}[baseline=(gamma.base)]
    \path node[inner sep=0pt] (gamma) {$\gamma$};
    \path (gamma) ++(40pt,0pt) node[Sdot] (S) {};
    \draw (gamma) -- ++(-10pt,-15pt);
    \node[draw,dotted,thick,inner sep=2pt,circle,fit =(gamma)] (C1) {};        
    \node[draw,dotted,thick,inner sep=2pt,circle,fit =(S)] (C2) {};        
    \draw[dashed] (C1) .. controls +(10pt,20pt) and +(-10pt,20pt) .. coordinate (a) (C2);
    \path (C1.north) node[anchor=south] {$\scriptstyle q_1$};
    \path (C2.north) node[anchor=south] {$\scriptstyle q_1$};
    \node[draw,dotted,thick,inner sep=0pt,ellipse,fit =(C1) (C2) (a)] (C) {};        
    \path (C.north) ++(-1pt,3pt) node[anchor= west] {$\scriptstyle q_a$};
    \draw[dashed] (C.north) -- ++(0,10pt);
  \end{tikzpicture}
  \, + \,
    \begin{tikzpicture}[baseline=(gamma.base)]
    \path node[inner sep=0pt] (gamma) {$\gamma$};
    \path (gamma) ++(40pt,0pt) node[inner sep=1pt,circle,draw] (b) {$B$};
    \draw (gamma) -- ++(-10pt,-15pt);
    \node[draw,dotted,thick,inner sep=2pt,circle,fit =(gamma)] (C1) {};        
    \draw[dashed] (C1) .. controls +(10pt,20pt) and +(-10pt,20pt) .. coordinate (a) (b);
    \path (C1.north) node[anchor=south] {$\scriptstyle q_1$};
    \node[draw,dotted,thick,inner sep=0pt,ellipse,fit =(C1) (b) (a)] (C) {};        
    \path (C.north) ++(-1pt,3pt) node[anchor= west] {$\scriptstyle q_a$};
    \draw[dashed] (C.north) -- ++(0,10pt);
  \end{tikzpicture}
  \notag \displaybreak[0] \\ & =\, 
  \begin{tikzpicture}[baseline=(gamma.base)]
    \path node[inner sep=0pt] (gamma) {$\gamma$};
    \path (gamma) ++(40pt,0pt) node[Sdot] (S) {};
    \draw (gamma) -- ++(-10pt,-15pt);
    \node[draw,dotted,thick,inner sep=2pt,circle,fit =(gamma)] (C1) {};        
    \node[draw,dotted,thick,inner sep=2pt,circle,fit =(S)] (C2) {};        
    \draw[dashed] (C1) .. controls +(10pt,20pt) and +(-10pt,20pt) .. (C2);
    \path (C1.north east) node[anchor=west] {$\scriptstyle q_1$};
    \path (C2.north) node[anchor=south] {$\scriptstyle q_1$};
    \node[draw,dotted,thick,inner sep=3pt,ellipse,fit =(gamma)] (C) {};        
    \path (C.north) ++(1pt,3pt) node[anchor= east] {$\scriptstyle q_a$};
    \draw[dashed] (C.north) -- ++(0,20pt);
  \end{tikzpicture}
  \, + \,
  \begin{tikzpicture}[baseline=(gamma.base)]
    \path node[inner sep=0pt] (gamma) {$\gamma$};
    \path (gamma) ++(40pt,0pt) node[Sdot] (S) {};
    \draw (gamma) -- ++(-10pt,-15pt);
    \node[draw,dotted,thick,inner sep=2pt,circle,fit =(gamma)] (C1) {};        
    \node[draw,dotted,thick,inner sep=2pt,circle,fit =(S)] (C2) {};        
    \draw[dashed] (C1) .. controls +(10pt,20pt) and +(-10pt,20pt) .. (C2);
    \path (C1.north) node[anchor=south] {$\scriptstyle q_1$};
    \path (C2.north west) node[anchor=east] {$\scriptstyle q_1$};
    \node[draw,dotted,thick,inner sep=3pt,ellipse,fit =(S)] (C) {};        
    \path (C.north) ++(-1pt,3pt) node[anchor= west] {$\scriptstyle q_a$};
    \draw[dashed] (C.north) -- ++(0,20pt);
  \end{tikzpicture}
  \, + \,
  \begin{tikzpicture}[baseline=(gamma.base)]
    \path node[inner sep=0pt] (gamma) {$\gamma$};
    \path (gamma) ++(40pt,0pt) node[inner sep=1pt,circle,draw] (b) {$B$};
    \draw (gamma) -- ++(-10pt,-15pt);
    \node[draw,dotted,thick,inner sep=2pt,circle,fit =(gamma)] (C1) {};        
    \draw[dashed] (C1) .. controls +(10pt,20pt) and +(-10pt,20pt) .. (b);
    \path (C1.north east) node[anchor=west] {$\scriptstyle q_1$};
    \node[draw,dotted,thick,inner sep=3pt,ellipse,fit =(gamma)] (C) {};        
    \path (C.north) ++(1pt,3pt) node[anchor= east] {$\scriptstyle q_a$};
    \draw[dashed] (C.north) -- ++(0,20pt);
  \end{tikzpicture}
  \, + \,
  \begin{tikzpicture}[baseline=(gamma.base)]
    \path node[inner sep=0pt] (gamma) {$\gamma$};
    \path (gamma) ++(40pt,0pt) node[inner sep=1pt,circle,draw] (b) {$B$};
    \draw (gamma) -- ++(-10pt,-15pt);
    \node[draw,dotted,thick,inner sep=2pt,circle,fit =(gamma)] (C1) {};        
    \draw[dashed] (C1) .. controls +(10pt,20pt) and +(-10pt,20pt) .. (b);
    \path (C1.north) node[anchor=south] {$\scriptstyle q_1$};
    \node[draw,dotted,thick,inner sep=0pt,ellipse,fit =(b)] (C) {};        
    \path (C.north) ++(-1pt,3pt) node[anchor= west] {$\scriptstyle q_a$};
    \draw[dashed] (C.north) -- ++(0,20pt);
  \end{tikzpicture}
  \notag \displaybreak[0] \\ & =\, 
  \begin{tikzpicture}[baseline=(gamma.base)]
    \path node[inner sep=0pt] (gamma) {$\gamma$};
    \path (gamma) ++(40pt,0pt) node[Sdot] (S) {};
    \node[draw,dotted,thick,inner sep=2pt,circle,fit =(gamma)] (C1) {};        
    \node[draw,dotted,thick,inner sep=2pt,circle,fit =(S)] (C2) {};        
    \draw[dashed] (C1) .. controls +(10pt,20pt) and +(-10pt,20pt) .. (C2);
    \path (C1.north) node[anchor=south] {$\scriptstyle q_1$};
    \path (C2.north) node[anchor=south] {$\scriptstyle q_1$};
    \path (gamma) ++(-18pt,-15pt) node[dot] (O) {} ++(0pt,20pt) node[inner sep=2pt] (gamma1) {$\gamma$};
    \draw (gamma) -- (O) -- (gamma1);
    \draw (O) .. controls +(-10pt,15pt) and +(0,15pt) .. ++(-20pt,-5pt);
    \node[draw,dotted,thick,inner sep=0pt,ellipse,fit =(gamma1)] (C) {};        
    \path (C.north) ++(1pt,3pt) node[anchor= east] {$\scriptstyle q_a$};
    \draw[dashed] (C.north) -- ++(0,15pt);
  \end{tikzpicture}
  \, + \,
  \begin{tikzpicture}[baseline=(gamma.base)]
    \path node[inner sep=0pt] (gamma) {$\gamma$};
    \path (gamma) ++(40pt,0pt) node[inner sep=1pt,circle,draw] (b) {$B$};
    \node[draw,dotted,thick,inner sep=2pt,circle,fit =(gamma)] (C1) {};        
    \draw[dashed] (C1) .. controls +(10pt,20pt) and +(-10pt,20pt) .. (b);
    \path (C1.north) node[anchor=south] {$\scriptstyle q_1$};
    \path (gamma) ++(-18pt,-15pt) node[dot] (O) {} ++(0pt,20pt) node[inner sep=2pt] (gamma1) {$\gamma$};
    \draw (gamma) -- (O) -- (gamma1);
    \draw (O) .. controls +(-10pt,15pt) and +(0,15pt) .. ++(-20pt,-5pt);
    \node[draw,dotted,thick,inner sep=0pt,ellipse,fit =(gamma1)] (C) {};        
    \path (C.north) ++(1pt,3pt) node[anchor= east] {$\scriptstyle q_a$};
    \draw[dashed] (C.north) -- ++(0,15pt);
  \end{tikzpicture}
  \, + \,
  \begin{tikzpicture}[baseline=(gamma.base)]
    \path node[inner sep=0pt] (gamma) {$\gamma$};
    \path (gamma) ++(40pt,0pt) node[inner sep=0pt] (gamma1) {$\gamma$} ++(10pt,-15pt) node[dot] (O) {} ++(10pt,15pt) node[inner sep=0pt] (gamma2) {$\gamma$};
    \draw (gamma1) -- (O) -- (gamma2);
    \draw (gamma) -- ++(-10pt,-20pt);
    \node[draw,dotted,thick,inner sep=2pt,circle,fit =(gamma)] (C1) {};        
    \node[draw,dotted,thick,inner sep=2pt,circle,fit =(gamma1)] (C2) {};        
    \draw[dashed] (C1) .. controls +(10pt,20pt) and +(-10pt,20pt) .. (C2);
    \path (C1.north) node[anchor=south] {$\scriptstyle q_1$};
    \path (C2.north) node[anchor=south] {$\scriptstyle q_1$};
    \node[draw,dotted,thick,inner sep=2pt,ellipse,fit =(gamma2)] (C) {};        
    \path (C.north) ++(-1pt,3pt) node[anchor= west] {$\scriptstyle q_a$};
    \draw[dashed] (C.north) -- ++(0,20pt);
  \end{tikzpicture}
  \, + \,
  \begin{tikzpicture}[baseline=(gamma.base)]
    \path node[inner sep=0pt] (gamma) {$\gamma$};
    \path (gamma) ++(40pt,0pt) node[inner sep=1pt,circle,draw] (b) {$B$};
    \draw (gamma) -- ++(-10pt,-20pt);
    \node[draw,dotted,thick,inner sep=2pt,circle,fit =(gamma)] (C1) {};        
    \draw[dashed] (C1) .. controls +(10pt,20pt) and +(-10pt,20pt) .. (b);
    \path (C1.north) node[anchor=south] {$\scriptstyle q_1$};
    \node[draw,dotted,thick,inner sep=0pt,ellipse,fit =(b)] (C) {};        
    \path (C.north) ++(-1pt,3pt) node[anchor= west] {$\scriptstyle q_a$};
    \draw[dashed] (C.north) -- ++(0,20pt);
  \end{tikzpicture}
  \notag \displaybreak[0] \\ & =\, 
  \begin{tikzpicture}[baseline=(gamma.base)]
    \path node[inner sep=0pt] (gamma) {$\gamma$};
    \node[draw,dotted,thick,inner sep=2pt,circle,fit =(gamma)] (C1) {};        
    \path (C1.north east) node[anchor=west] {$\scriptstyle t$};
    \path (gamma) ++(-18pt,-15pt) node[dot] (O) {} ++(0pt,20pt) node[inner sep=2pt] (gamma1) {$\gamma$};
    \draw (gamma) -- (O) -- (gamma1);
    \draw (O) .. controls +(-10pt,15pt) and +(0,15pt) .. ++(-20pt,-5pt);
    \node[draw,dotted,thick,inner sep=0pt,ellipse,fit =(gamma1)] (C) {};        
    \path (C.north) ++(1pt,3pt) node[anchor= east] {$\scriptstyle q_a$};
    \draw[dashed] (C.north) -- ++(0,15pt);
  \end{tikzpicture}
  \, + \,
  \begin{tikzpicture}[baseline=(gamma.base)]
    \path node[inner sep=0pt] (gamma) {$\gamma$};
    \path (gamma) ++(40pt,0pt) node[inner sep=0pt] (gamma1) {$\gamma$} ++(10pt,-15pt) node[dot] (O) {} ++(10pt,15pt) node[inner sep=0pt] (gamma2) {$\gamma$};
    \draw (gamma1) -- (O) -- (gamma2);
    \draw (gamma) -- ++(-10pt,-20pt);
    \node[draw,dotted,thick,inner sep=2pt,circle,fit =(gamma)] (C1) {};        
    \node[draw,dotted,thick,inner sep=2pt,circle,fit =(gamma1)] (C2) {};        
    \draw[dashed] (C1) .. controls +(10pt,20pt) and +(-10pt,20pt) .. (C2);
    \path (C1.north) node[anchor=south] {$\scriptstyle q_1$};
    \path (C2.north) node[anchor=south] {$\scriptstyle q_1$};
    \node[draw,dotted,thick,inner sep=2pt,ellipse,fit =(gamma2)] (C) {};        
    \path (C.north) ++(-1pt,3pt) node[anchor= west] {$\scriptstyle q_a$};
    \draw[dashed] (C.north) -- ++(0,20pt);
  \end{tikzpicture}
  \, + \,
  \begin{tikzpicture}[baseline=(gamma.base)]
    \path node[inner sep=0pt] (gamma) {$\gamma$};
    \path (gamma) ++(40pt,0pt) node[inner sep=1pt,circle,draw] (b) {$B$};
    \draw (gamma) -- ++(-10pt,-20pt);
    \node[draw,dotted,thick,inner sep=2pt,circle,fit =(gamma)] (C1) {};        
    \draw[dashed] (C1) .. controls +(10pt,20pt) and +(-10pt,20pt) .. (b);
    \path (C1.north) node[anchor=south] {$\scriptstyle q_1$};
    \node[draw,dotted,thick,inner sep=0pt,ellipse,fit =(b)] (C) {};        
    \path (C.north) ++(-1pt,3pt) node[anchor= west] {$\scriptstyle q_a$};
    \draw[dashed] (C.north) -- ++(0,20pt);
  \end{tikzpicture}
  \label{dgammadTQ}
\end{align}
We remark that $\tikz[baseline=(b.base)] { \node[draw,circle,inner sep=1pt] (b) {$B$}; \node[draw,dotted,thick,inner sep=0pt,ellipse,fit =(b)] (C) {}; \draw[dashed] (b.north west) -- ++(-5pt,10pt); \draw[dashed] (C.north east) -- ++(5pt,10pt); \path (b.north east) node[anchor=west] {$\scriptstyle q_0$}; } \!\! = \frac{\partial B(q_1)}{\partial q_0} = 0$.

We turn now to \tikz[baseline=(S.base)] \path node[draw,ellipse,inner sep=1pt] (S) {$(\partial^2[-S])^{-1}$};.  Recall \cref{HJ}.  Then:
\begin{align}
  \begin{tikzpicture}[baseline=(S.south)]
    \path node[Sdot] (S) {} ++(0,0pt) coordinate (T);
    \path (S) node[draw,dotted,thick,inner sep=5pt,circle] (C1) {};        
    \path (S) node[draw,dotted,thick,inner sep=6pt,circle] (C2) {};        
    \path (S) node[draw,dotted,thick,inner sep=7pt,circle] (C) {};        
    \path (S) +(-15pt,25pt) coordinate (t1) +(15pt,25pt) coordinate (t2);
    \draw[dashed] (C2.north west) .. controls +(-5pt,5pt) and +(0pt,-10pt) .. (t1);
    \draw[dashed] (C2.north east) .. controls +(5pt,5pt) and +(0pt,-10pt) .. (t2);
    \path (C2.north west) node[anchor=east] {$\scriptstyle q_0$};
    \path (C2.north east) node[anchor=west] {$\scriptstyle q_1$};
    \path (C.south west) node[anchor=north] {$\scriptstyle t$};
  \end{tikzpicture}
  \, & = \,
  \left(\frac12\right)
   \begin{tikzpicture}[baseline=(S1.south)]
    \path node[Sdot] (S1) {};
    \path (S1) ++(40pt,0pt) node[Sdot] (S2) {};
    \node[draw,dotted,thick,inner sep=2pt,circle,fit =(S1)] (C1) {};        
    \node[draw,dotted,thick,inner sep=2pt,circle,fit =(S2)] (C2) {};        
    \draw[dashed] (C1) .. controls +(10pt,20pt) and +(-10pt,20pt) .. coordinate (a) (C2);
    \path (C1.north) node[anchor=south] {$\scriptstyle q_1$};
    \path (C2.north) node[anchor=south] {$\scriptstyle q_1$};
    \node[draw,dotted,thick,inner sep=0pt,ellipse,fit = (C1) (C2) (a)] (Ci) {};
    \node[draw,dotted,thick,inner sep=1pt,ellipse,fit = (C1) (C2) (a)] (C) {};
    \draw[dashed] (C.north west) -- ++(0pt,15pt);
    \draw[dashed] (C.north east) -- ++(0pt,15pt);
    \path (C.north west) node[anchor=east] {$\scriptstyle q_0$};
    \path (C.north east) node[anchor=west] {$\scriptstyle q_1$};
  \end{tikzpicture}
  \, + \,
   \begin{tikzpicture}[baseline=(S1.south)]
    \path node[Sdot] (S1) {};
    \path (S1) ++(40pt,0pt) node[inner sep=1pt,circle,draw] (b) {$B$};
    \node[draw,dotted,thick,inner sep=2pt,circle,fit =(S1)] (C1) {};        
    \draw[dashed] (C1) .. controls +(10pt,20pt) and +(-10pt,20pt) .. coordinate (a) (b);
    \path (C1.north) node[anchor=south] {$\scriptstyle q_1$};
    \node[draw,dotted,thick,inner sep=0pt,ellipse,fit = (C1) (b) (a)] (Ci) {};
    \node[draw,dotted,thick,inner sep=1pt,ellipse,fit = (C1) (b) (a)] (C) {};
    \draw[dashed] (C.north west) -- ++(0pt,15pt);
    \draw[dashed] (C.north east) -- ++(0pt,15pt);
    \path (C.north west) node[anchor=east] {$\scriptstyle q_0$};
    \path (C.north east) node[anchor=west] {$\scriptstyle q_1$};
  \end{tikzpicture}
  \, + \,
  \left(\frac12\right)
   \begin{tikzpicture}[baseline=(b1.base)]
    \path node[inner sep=1pt,circle,draw] (b1) {$B$};
    \path (b1) ++(40pt,0pt) node[inner sep=1pt,circle,draw] (b) {$B$};
    \draw[dashed] (b1) .. controls +(10pt,20pt) and +(-10pt,20pt) .. coordinate (a) (b);
    \node[draw,dotted,thick,inner sep=0pt,ellipse,fit = (b1) (b) (a)] (Ci) {};
    \node[draw,dotted,thick,inner sep=1pt,ellipse,fit = (b1) (b) (a)] (C) {};
    \draw[dashed] (C.north west) -- ++(0pt,15pt);
    \draw[dashed] (C.north east) -- ++(0pt,15pt);
    \path (C.north west) node[anchor=east] {$\scriptstyle q_0$};
    \path (C.north east) node[anchor=west] {$\scriptstyle q_1$};
  \end{tikzpicture}  
  \, + \,
  \tikz[baseline=(c.base)] {\path node[draw,inner sep=1pt,circle] (c) {$C$} node[circle,draw,dotted,thick,inner sep=0pt,fit=(c)] {} node[circle,draw,dotted,thick,inner sep=1pt,fit=(c)] (C2) {};      \path (c) +(-15pt,25pt) coordinate (t1) +(15pt,25pt) coordinate (t2);
    \draw[dashed] (C2.north west) .. controls +(-5pt,5pt) and +(0pt,-10pt) .. (t1);
    \draw[dashed] (C2.north east) .. controls +(5pt,5pt) and +(0pt,-10pt) .. (t2);
    \path (C2.north west) node[anchor=east] {$\scriptstyle q_0$};
    \path (C2.north east) node[anchor=west] {$\scriptstyle q_1$}; }
  \notag \\ & = \,
 \label{dSdQQT}
\end{align}
In the second line above we used the product rule and dropped any terms involving derivatives with respect to $q_0$ of functions only of $q_1$.  In the third line we used \cref{dSdQQ,dSdQQQ,dgammadt}.

By the quotient rule:
\begin{multline}
  %
  \label{dSdQQidT}
\end{multline}
Thus, we can differentiate the Green's function by recalling \cref{greeneqn2} and using \cref{dgammadTQ,dSdQQidT}.  Many terms cancel:
\begin{align}
  %
 \label{greendifferentiationpart}
\end{align}
Finally, consider the integrals implicit in the trivalent vertices in \cref{greendifferentiationpart}.  Write $\rho$ for the variable of integration for each, and consider the sum as a single integral $\int_{\rho = 0}^t$.  When $\rho \leq \varsigma \leq \tau$, the second and third terms exactly cancel; when $\varsigma \leq \rho \leq \tau$, the three terms agree up to sign; and when $\varsigma \leq \tau \leq \rho$, the first two terms cancel.  (When $\tau \leq \varsigma$, the contribution to the total expression is from the part we have abbreviate ``$(\varsigma \tofrom \tau)$'', and the analysis is similar.)  But consider breaking the integral

as a similar sum of contributions from $\rho \leq \varsigma \leq \tau$, etc.  \Cref{greeneqn2} implies that the contributions from each part exactly match the integrals in \cref{greendifferentiationpart}.  This proves:
\begin{equation} \label{dGdT}
  %
\end{equation}

Given \cref{dSdQQQ,dSdQQT} and the well-known formula for the logarithmic derivative of a determinant, we also have:
\begin{align} \label{dDETdQ}
  \frac{\partial}{\partial q_a} \left[ \log \left| \det \frac{\partial^2[-S]}{\partial q_0\partial q_1}\right|\right] & = 
  %
\end{align}
We conclude this section by simplifying the right-hand side of \cref{dDETdQ} and  the first term on the right-hand side of \cref{dDETdT}.  We first expand the \tikz \node[dot] {};-vertex:
\begin{equation*}
  %
\end{equation*}
But the integrals implicit in the first two summands above are obviously symmetric under switching the two ``$\gamma$'' legs.  Recalling \cref{greeneqn2}:
\begin{equation*}
  %
  
\end{equation*}
All together, we have the following simplification:
\begin{equation} \label{claimoneloop}
  %
  
\end{equation}

\subsection{The cancellations}\label{cancellations}

In the previous sections, we evaluated the derivatives of every component of $V_\gamma$.  In this section, we check \cref{SEV}, which in the graphical notation reads:\\[-1pc]
\begin{equation} \label{SEVdiagram}
\tikz[baseline=(V.base)] \node[inner sep=1pt,draw,dotted,thick,circle] (V) {$V_\gamma$} (V.north) ++(1pt,3pt) node[anchor= east] {$\scriptstyle t$}; \, =
  \,
  \frac12 \, 
   %
  
  \, + \,
  \tikz[baseline=(c.base)] \path node[draw,inner sep=1pt,ellipse] (c) {$C$};  
\end{equation}
We have adopted the following convention: the dashed lines (although not the dotted circles) count as edges and the new nodes count as vertices for the purposes of computing Euler characteristic, etc., and in equations with connected diagrams with different numbers of loops, we multiply each diagram $\Gamma$ by $(i\hbar)^{\lambda(\Gamma)}$.

Recalling that $V_\gamma = \tikz \node[Sdot] {}; \,+ \bigl( \frac{i\hbar}2\bigr) \log \bigl| \det \frac{\partial^2[-S_\gamma]}{\partial q_0\partial q_1}\bigr| + \sum\{\text{connected diagrams}\}/\{\text{symmetry}\}$, the calculations from the previous sections give:
\begin{equation} \label{cancelation2}
  \tikz[baseline=(V.base)] \node[inner sep=1pt,draw,dotted,thick,circle] (V) {$V_\gamma$} (V.north) ++(1pt,3pt) node[anchor= east] {$\scriptstyle t$}; \, 
   =   
  \tikz[baseline=(S.south)]{\node[Sdot] (S) {}; \node (C) [draw,circle,dotted,thick,inner sep=2pt, fit=(S)] {}; \path (C.north west) node[anchor=south] {$\scriptstyle t$};}
  \, +
  \frac12\,
  \begin{tikzpicture}[baseline=(gamma1.base)]
    \node[dot] (O) {};
    \path (O) ++(-15pt,10pt) node[inner sep=0pt] (gamma1) {$\gamma$}; 
    \path (O) ++(15pt,10pt) node[inner sep=0pt] (gamma2) {$\gamma$}; 
    \path (gamma1) node (C1) [draw,circle,dotted,thick,inner sep=5pt] {};
    \path (gamma2) node (C2) [draw,circle,dotted,thick,inner sep=5pt] {};
    \draw[dashed] (C1) .. controls +(10pt,15pt) and +(-10pt,15pt) .. (C2);
    \path (C1.north) node[anchor= south]{$\scriptstyle q_1$};
    \path (C2.north) node[anchor= south]{$\scriptstyle q_1$};
    \draw (O) -- (gamma1); \draw (O) -- (gamma2);
  \end{tikzpicture}
  \,+ \frac12 \!\!\!
  \begin{tikzpicture}[baseline=(B.base)]
    \path node[draw,circle,inner sep=1pt] (b) {$B$};
    \node[ellipse,draw,inner sep=1pt,dotted,thick,fit=(b)] (C) {};
    \draw[dashed] (b.north west) .. controls +(-20pt,20pt) and +(15pt,15pt) .. (C.north east);
    \path (C.north east) ++(-2pt,0pt) node[anchor=west] {$\scriptstyle q_1$};
  \end{tikzpicture}
  \!\!\! +
  \sum_{\substack{ \Gamma \text{ connected} \\ \text{with one  \tikz[baseline=(gamma.base)] \path node[inner sep=0pt] (gamma) {$\scriptstyle \gamma$} node[draw,dotted,thick,circle,inner sep=1pt,fit=(gamma)] (C) {} (C.north) ++(1pt,2pt) node[anchor= east] {$\scriptscriptstyle t$}; or}
  \\
  \text{one  \tikz[baseline=(gamma1.base)]{ \path node[inner sep=0pt] (gamma1) {$\scriptstyle \gamma$} node[draw,dotted,thick,circle,inner sep=1pt,fit=(gamma1)] (C1) {} (C1.north) ++(1pt,2pt) node[anchor= east] {$\scriptscriptstyle q_1$};  \path (gamma1) ++(25pt,0pt) node[inner sep=0pt] (gamma2) {$\scriptstyle \gamma$} node[draw,dotted,thick,circle,inner sep=1pt,fit=(gamma2)] (C2) {} (C2.north) ++(-1pt,2pt) node[anchor= west] {$\scriptscriptstyle q_1$}; \draw[dashed] (C1) -- (C2);}}
  }}
  \frac{(i\hbar)^{\lambda(\Gamma)}\Ff_\gamma(\Gamma)}{\left| \Aut\Gamma\right|}
\end{equation}
In the sum, it is implied that all other vertices are usual \tikz\node[dot] {};-vertices and are of valence $n\geq 3$.  The sum includes the diagram   \!\begin{tikzpicture}[baseline=(X)]
    \path node[dot] (O) {} ++(0pt,5pt) coordinate (X);
    \draw (O) .. controls +(15pt,15pt) and +(0pt,17pt) .. (O);
    \path (O) ++(-10pt,10pt)node[inner sep=0pt] (gamma) {$\gamma$};
    \node[draw,dotted,thick,circle,inner sep=1pt,fit=(gamma)] (C) {} (C.west) ++(-2pt,0pt) node[anchor=south] {$\scriptstyle t$};
    \draw (O) -- (gamma);
  \end{tikzpicture}\!\!
 coming from $\frac{\partial}{\partial t}\log \bigl| \det \frac{\partial^2[-S_\gamma]}{\partial q_0\partial q_1}\bigr|$.  Otherwise, each diagram with a \!\!\tikz[baseline=(gamma.base)] \path node[inner sep=0pt] (gamma) {$\gamma$} node[draw,dotted,thick,circle,inner sep=1pt,fit=(gamma)] (C) {} (C.west) ++(-2pt,0pt) node[anchor=south] {$\scriptstyle t$}; comes from the diagram formed by removing it (if the \!\!\tikz[baseline=(gamma.base)] \path node[inner sep=0pt] (gamma) {$\gamma$} node[draw,dotted,thick,circle,inner sep=1pt,fit=(gamma)] (C) {} (C.west) ++(-2pt,0pt) node[anchor=south] {$\scriptstyle t$}; attaches to a trivalent vertex, its removal leaves just a solid edge), and the combinatorics are as in \cref{productrule}: if there are $n$ equivalent ways to attach a \!\!\tikz[baseline=(gamma.base)] \path node[inner sep=0pt] (gamma) {$\gamma$} node[draw,dotted,thick,circle,inner sep=1pt,fit=(gamma)] (C) {} (C.west) ++(-2pt,0pt) node[anchor=south] {$\scriptstyle t$}; to a diagram $\Gamma$, then $\bigl| \Aut\, \tikz[baseline=(G.base)]{ \path node[rectangle,draw,inner sep=1pt] (G) {$\Gamma$}; \path (G.center)++(20pt,0pt) node[anchor=center,inner sep=0pt] (gamma) {$\gamma$} node[draw,dotted,thick,circle,inner sep=1pt,fit=(gamma)] (C) {} (C.west) ++(-1pt,0pt) node[anchor=south] {$\scriptstyle t$}; \draw (G) -- (gamma);} \bigr| = \frac1n \left| \Aut\Gamma\right|$.  The second sum occurs similarly, with each \!\tikz[baseline=(gamma1.base)]{ \path node[inner sep=0pt] (gamma1) {$\gamma$} node[draw,dotted,thick,circle,inner sep=1pt,fit=(gamma1)] (C1) {} (C1.west) ++(-3pt,0pt) node[anchor=south] {$\scriptstyle q_1$};  \path (gamma1) ++(25pt,0pt) node[inner sep=0pt] (gamma2) {$\gamma$} node[draw,dotted,thick,circle,inner sep=1pt,fit=(gamma2)] (C2) {}  (C2.east) ++(3pt,0pt) node[anchor=south] {$\scriptstyle q_1$}; \draw[dashed] (C1) -- (C2);}\! replacing a Green's function, as per the second term in \cref{dGdT}.

Similarly, we have:
\begin{equation}
  \tikz[baseline=(V.base)] \node[inner sep=1pt,draw,dotted,thick,circle] (V) {$V_\gamma$} (V.north) ++(1pt,3pt) node[anchor= east] {$\scriptstyle q_1$} (V.north) [draw,dashed] -- ++(0pt,15pt);
   =
  \tikz[baseline=(S.south)]{\node[Sdot] (S) {}; \node (C) [draw,circle,dotted,thick,inner sep=2pt, fit=(S)] {}; \path (C.north) ++(1pt,3pt) node[anchor= east] {$\scriptstyle q_1$}; \draw[dashed] (C.north) -- ++(0pt,15pt);} \,+
  \sum_{\substack{ \Gamma \text{ connected} \\ \text{with one  \tikz[baseline=(gamma.base)] \path node[inner sep=0pt] (gamma) {$\scriptstyle \gamma$} node[draw,dotted,thick,circle,inner sep=1pt,fit=(gamma)] (C) {} (C.north) ++(1pt,2pt) node[anchor= east] {$\scriptscriptstyle q_1$};}}}
  \frac{(i\hbar)^{\lambda(\Gamma)}\Ff_\gamma(\Gamma)}{\left| \Aut\Gamma\right|}
\end{equation}\\[-1pc]
Then $\frac12 \!\!\!
  \begin{tikzpicture}[baseline=(V.base)]
    \path node[draw,dotted,thick,circle,inner sep=1pt] (V) {$V_\gamma$};
    \node[ellipse,draw,inner sep=-1pt,dotted,thick,fit=(V)] (C) {};
    \draw[dashed] (C.north west) .. controls +(-15pt,15pt) and +(15pt,15pt) .. (C.north east);
    \path (C.north east) ++(-2pt,0pt) node[anchor=west] {$\scriptstyle q_1$};
    \path (C.north west) ++(2pt,0pt) node[anchor=east] {$\scriptstyle q_1$};
  \end{tikzpicture}
  \!\!\! =    \frac12\,
  \begin{tikzpicture}[baseline=(gamma1.base)]
    \node[dot] (O) {};
    \path (O) ++(-15pt,10pt) node[inner sep=0pt] (gamma1) {$\gamma$}; 
    \path (O) ++(15pt,10pt) node[inner sep=0pt] (gamma2) {$\gamma$}; 
    \path (gamma1) node (C1) [draw,circle,dotted,thick,inner sep=5pt] {};
    \path (gamma2) node (C2) [draw,circle,dotted,thick,inner sep=5pt] {};
    \draw[dashed] (C1) .. controls +(10pt,15pt) and +(-10pt,15pt) .. (C2);
    \path (C1.north) node[anchor= south]{$\scriptstyle q_1$};
    \path (C2.north) node[anchor= south]{$\scriptstyle q_1$};
    \draw (O) -- (gamma1); \draw (O) -- (gamma2);
  \end{tikzpicture}
  \,+ $ a sum of diagrams each with one \!\tikz[baseline=(gamma1.base)]{ \path node[inner sep=0pt] (gamma1) {$\gamma$} node[draw,dotted,thick,circle,inner sep=1pt,fit=(gamma1)] (C1) {} (C1.west) ++(-3pt,0pt) node[anchor=south] {$\scriptstyle q_1$};  \path (gamma1) ++(25pt,0pt) node[inner sep=0pt] (gamma2) {$\gamma$} node[draw,dotted,thick,circle,inner sep=1pt,fit=(gamma2)] (C2) {}  (C2.east) ++(3pt,0pt) node[anchor=south] {$\scriptstyle q_1$}; \draw[dashed] (C1) -- (C2);}\!, such that its removal yields a diagram that is still connected.
  
The diagrams in  $\frac12 
\tikz[baseline=(gamma1.base)]{ \path node[inner sep=0pt] (gamma1) {$V_\gamma$} node[draw,dotted,thick,circle,inner sep=1pt,fit=(gamma1)] (C1) {} (C1.west) ++(-3pt,0pt) node[anchor=south] {$\scriptstyle q_1$};  \path (gamma1) ++(35pt,0pt) node[inner sep=0pt] (gamma2) {$V_\gamma$} node[draw,dotted,thick,circle,inner sep=1pt,fit=(gamma2)] (C2) {}  (C2.east) ++(3pt,0pt) node[anchor=south] {$\scriptstyle q_1$}; \draw[dashed] (C1) -- (C2);}\!
$ come in two types.  Other than the simply-connected diagram $\frac12 
\tikz[baseline=(gamma1.base)]{ \path node[Sdot] (gamma1) {} node[draw,dotted,thick,circle,inner sep=2pt,fit=(gamma1)] (C1) {} (C1.west) ++(-3pt,0pt) node[anchor=south] {$\scriptstyle q_1$};  \path (gamma1) ++(25pt,0pt) node[Sdot] (gamma2) {} node[draw,dotted,thick,circle,inner sep=2pt,fit=(gamma2)] (C2) {}  (C2.east) ++(3pt,0pt) node[anchor=south] {$\scriptstyle q_1$}; \draw[dashed] (C1) -- (C2);}\!
$, there are diagrams where the dashed line connects a complicated part to a \tikz[baseline=(gamma1.base)] \path node[Sdot] (gamma1) {} node[draw,dotted,thick,circle,inner sep=2pt,fit=(gamma1)] (C1) {} (C1.west) ++(-3pt,0pt) node[anchor=south] {$\scriptstyle q_1$};, and diagrams where the dashed line connects two non-simply-connected diagrams, which are not connected otherwise.  But recall \cref{dgammadt}; then:
\begin{multline} \label{cancelation1}
  \frac12 \, 
   \begin{tikzpicture}[baseline=(V1.base)]
    \path node[inner sep=1pt,draw,dotted,thick,circle] (V1) {$V_\gamma$};
    \path (V1.base) ++(40pt,0pt) node[anchor=base,inner sep=1pt,draw,dotted,thick,circle] (V2) {$V_\gamma$};
    \draw[dashed] (V1) .. controls +(10pt,20pt) and +(-10pt,20pt) .. (V2);
    \path (V1.north) node[anchor=south] {$\scriptstyle q_1$};
    \path (V2.north) node[anchor=south] {$\scriptstyle q_1$};
  \end{tikzpicture}
  \, + \,
   \begin{tikzpicture}[baseline=(V1.base)]
    \path node[inner sep=1pt,draw,dotted,thick,circle] (V1) {$V_\gamma$} +(0,7pt);
    \path (V1.base) ++(40pt,0pt) node[anchor=base,inner sep=1pt,circle,draw] (b) {$B$};
    \draw[dashed] (V1) .. controls +(10pt,20pt) and +(-10pt,20pt) .. (b);
    \path (V1.north) node[anchor=south] {$\scriptstyle q_1$};
  \end{tikzpicture}
  \, + \frac12 \!\!\!
  \begin{tikzpicture}[baseline=(V.base)]
    \path node[draw,dotted,thick,circle,inner sep=1pt] (V) {$V_\gamma$};
    \node[ellipse,draw,inner sep=-1pt,dotted,thick,fit=(V)] (C) {};
    \draw[dashed] (C.north west) .. controls +(-15pt,15pt) and +(15pt,15pt) .. (C.north east);
    \path (C.north east) ++(-2pt,0pt) node[anchor=west] {$\scriptstyle q_1$};
    \path (C.north west) ++(2pt,0pt) node[anchor=east] {$\scriptstyle q_1$};
  \end{tikzpicture}
  \!\!\!    = \\
  =
  \,
  \frac12 \, 
   \begin{tikzpicture}[baseline=(V1.base)]
    \path node[Sdot] (V1) {}; \node[draw,dotted,thick,inner sep=2pt,circle, fit=(V1)] (C1) {};
    \path (V1) ++(40pt,0pt) node[Sdot] (V2) {}; \node[draw,dotted,thick,inner sep=2pt,circle, fit=(V2)] (C2) {};
    \draw[dashed] (C1) .. controls +(10pt,20pt) and +(-10pt,20pt) .. (C2);
    \path (C1.north) node[anchor=south] {$\scriptstyle q_1$};
    \path (C2.north) node[anchor=south] {$\scriptstyle q_1$};
  \end{tikzpicture}
  \, + \,
   \begin{tikzpicture}[baseline=(V1.base)]
    \path node[Sdot] (V1) {}; \node[draw,dotted,thick,inner sep=2pt,circle, fit=(V1)] (C1) {};
    \path (V1.center) ++(40pt,0pt) node[anchor=center,inner sep=1pt,circle,draw] (b) {$B$};
    \draw[dashed] (C1) .. controls +(10pt,20pt) and +(-10pt,20pt) .. (b);
    \path (C1.north) node[anchor=south] {$\scriptstyle q_1$};
  \end{tikzpicture}
  \, +
\frac12\,
  \begin{tikzpicture}[baseline=(gamma1.base)]
    \node[dot] (O) {};
    \path (O) ++(-15pt,10pt) node[inner sep=0pt] (gamma1) {$\gamma$}; 
    \path (O) ++(15pt,10pt) node[inner sep=0pt] (gamma2) {$\gamma$}; 
    \path (gamma1) node (C1) [draw,circle,dotted,thick,inner sep=5pt] {};
    \path (gamma2) node (C2) [draw,circle,dotted,thick,inner sep=5pt] {};
    \draw[dashed] (C1) .. controls +(10pt,15pt) and +(-10pt,15pt) .. (C2);
    \path (C1.north) node[anchor= south]{$\scriptstyle q_1$};
    \path (C2.north) node[anchor= south]{$\scriptstyle q_1$};
    \draw (O) -- (gamma1); \draw (O) -- (gamma2);
  \end{tikzpicture}
  \,+
  \sum_{\substack{ \Gamma \text{ connected} \\ \text{with one  \tikz[baseline=(gamma.base)] \path node[inner sep=0pt] (gamma) {$\scriptstyle \gamma$} node[draw,dotted,thick,circle,inner sep=1pt,fit=(gamma)] (C) {} (C.north) ++(1pt,2pt) node[anchor= east] {$\scriptscriptstyle t$}; or}  
  \\
  \text{one  \tikz[baseline=(gamma1.base)]{ \path node[inner sep=0pt] (gamma1) {$\scriptstyle \gamma$} node[draw,dotted,thick,circle,inner sep=1pt,fit=(gamma1)] (C1) {} (C1.north) ++(1pt,2pt) node[anchor= east] {$\scriptscriptstyle q_1$};  \path (gamma1) ++(25pt,0pt) node[inner sep=0pt] (gamma2) {$\scriptstyle \gamma$} node[draw,dotted,thick,circle,inner sep=1pt,fit=(gamma2)] (C2) {} (C2.north) ++(-1pt,2pt) node[anchor= west] {$\scriptscriptstyle q_1$}; \draw[dashed] (C1) -- (C2);}}
  }}
  \frac{(i\hbar)^{\lambda(\Gamma)}\Ff_\gamma(\Gamma)}{\left| \Aut\Gamma\right|}
\end{multline}
But comparing \cref{cancelation1,cancelation2,SEVdiagram}, we see that the only remaining terms are equal by \cref{HJ}.  Thus we have proven \cref{MainThm}.

\Section{Proof of Theorem \texorpdfstring{\ref{IVPthm}}{2.3.3}} \label{IVPthmproof}

Recall that the pointwise convergence of distributions is defined by testing against compactly-supported functions.  For a chosen open neighborhood $\Oo \subseteq \RR^{2n+1}$, let $A$ be the set of all smooth families of classical paths with boundary values in $\Oo$, and fix $(t,q_1) \in \RR^{n+1}$.  Throughout this section, we will let $g: \RR^n \to \RR$ be smooth with compact support $K$ such that $\{t\} \times K \times \{q_1\} \subseteq \Oo$.  We will prove \cref{IVPthm} in three steps.  In \cref{step1}, we characterize those families $\gamma \in A$ such that $\int_K^\formal \exp\bigl( -(i\hbar)^{-1}V_\gamma(t,q_0,q_1)\bigr) \, g(q_0)\,dq_0$ is non-zero, and show that there are only finitely many such families.  In \cref{step2} we pick $\Oo_0 \subseteq \RR^n$ open with compact closure and show that for $\epsilon>0$ sufficiently small depending on $\Oo_0$, for $\Oo = (0,\epsilon) \times \Oo_0 \times \Oo_0$ there is a unique family of classical paths $\gamma$ with $\int_K^\formal \exp\bigl( -(i\hbar)^{-1}V_\gamma(t,q_0,q_1)\bigr) \, g(q_0)\,dq_0 \neq 0$.  In \cref{step3} we compute the limit as $t\to 0$ of this formal integral.

\subsection{On the support of formal integrals} \label{step1}

We wish to understand the formal integral:
\begin{equation}\label{IVPproofeqn1} \int^{\formal}_K \exp\bigl( (i\hbar)^{-1}V_\gamma(t,q_0,q_1)\bigr)\,g(q_0)\,dq_0 \end{equation}
We begin by splitting the exponent $V_\gamma = (-S_\gamma) + (V_\gamma + S_\gamma)$.  Then $V_\gamma + S_\gamma$ is valued in formal power series that begin in degree $\hbar$, so $\exp\bigl((i\hbar)^{-1}(V_\gamma + S_\gamma)\bigr)$ is a formal power series.  Thus the formal integral in \cref{IVPproofeqn1} is determined by the $\hbar\to 0$ asymptotics of integrals of the form
$ \int_K \exp\bigl( -(i\hbar)^{-1}S_\gamma(t,q_0,q_1)\bigr)\,g(q_0)\,dq_0 $.

Such integrals are conventionally studied by the method of stationary phase.  Indeed, by \cite[Lemma 3.13]{EZ2007}, if $\frac{\partial S_\gamma}{\partial q_0} \neq 0$ in $K$, then:
\begin{equation}
  \int_K \exp\bigl( -(i\hbar)^{-1}S_\gamma(t,q_0,q_1)\bigr)\,g(q_0)\,dq_0 = O(\hbar^\infty)\text{ as } \hbar \to 0
\end{equation}
By \cref{wellknown2}, $\frac{\partial S_\gamma}{\partial q_0}(t,q_0,q_1) = 0$ if and only if $\dot\gamma_{(t,q_0,q_1)}(0) = -B(q_0)$, where $\gamma_{(t,q_0,q_1)}$ is the member of the family of paths $\gamma$ that connects $q_0$ to $q_1$ in time $t$. 
 If the family $\gamma$ contains a path with initial values $(\dot\gamma(0),\gamma(0)) = (-B(q_0),q_0)$ for some $q_0 \in K$, then it does so in an open neighborhood of $q_0$, and $\gamma$ is determined on the connected component of $\{t\}\times K \times\{q_1\}$ containing $(t,q_0,q_1)$. 
 Thus, as $K$ is compact, there are only finitely many families $\gamma$ with $\frac{\partial S_\gamma}{\partial q_0}(t,q_0,q_1) = 0$ for some $q_0 \in K$, and so there are only finitely many paths $\gamma$ such that the formal integral in \cref{IVPproofeqn1} is non-zero as a formal expression in $\hbar$. 

\subsection{Short-duration classical paths can be almost geodesic} \label{step2}

For $\epsilon \in \RR$, consider the non-degenerate second-order ODE given by:
  \begin{equation} \label{EOMepsilon}
    0 = \ddot\gamma^i + \epsilon \left( \frac{\partial B_i}{\partial q^j}\bigl(\gamma\bigr) - \frac{\partial B_j}{\partial q^i}\bigl(\gamma\bigr) \right) \dot\gamma^j + \epsilon^2 \,\frac{\partial C}{\partial q^i}\bigl(\gamma\bigr)
  \end{equation}
Let $\varphi_\epsilon$ be the ``flow for duration $1$'' map, i.e.\ $\varphi_\epsilon(v,q) = \bigl(\dot\gamma_\epsilon(1),\gamma_\epsilon(1)\bigr)$ for $\gamma_\epsilon$ the unique solution to \cref{EOMepsilon} with $\bigl(\dot\gamma_\epsilon(0),\gamma_\epsilon(0)\bigr) = (v,q)$.  It is defined on an open neighborhood of the $0$ section in $\T\RR^n = \RR^{2n}$, it is smooth when it is defined, and it depends smoothly on $\epsilon$.  Let $\phi_\epsilon(v,q) = \bigl(q,\project \circ \varphi_\epsilon(v,q)\bigr)$, where $\project : \T\RR^n \to \RR^n$ is the natural projection.  Then $\phi_0$ is the isomorphism $(v,q) \mapsto (q,q+v)$.  So the partial function $\phi_{[-]}: \T\RR \times \RR \to \RR^{2n}$ is defined on an open neighborhood of $\T\RR^n \times \{0\}$.

Fix $\Oo_0 \subseteq \RR^n$ with compact closure $\overline{\Oo_0}$, and let $\Oo_1 \supseteq \overline{\Oo_0}$ be open with compact closure $\overline{\Oo_1}$.  Let $\Pp = \phi_0^{-1}\bigl(\overline{\Oo_1} \times \overline{\Oo_1}\bigr)$. It is a compact subset of $\T\RR^n$ containing $\{0\} \times \overline{\Oo_1}$.  Then we can find $\epsilon_0 > 0$ such that $\Pp \times (-\epsilon_0,\epsilon_0)$ is contained in the domain of $\phi_{[-]}$.  Let $(v,q) \in \Pp$ such that $q \in \Oo_1$.  For $\epsilon \in (-\epsilon_0,\epsilon_0)$, let $\gamma_\epsilon$ be the duration-$1$ solution to \cref{EOMepsilon} with initial conditions $\bigl(\dot\gamma_\epsilon(0),\gamma_\epsilon(0)\bigr) = (v,q)$.  Then $\gamma_0$ is nondegenerate, and the nondegeneracy condition depends smoothly on $\epsilon$.  Thus for each $(v,q)$ there is some number $0 < \epsilon_1(v,q) < \epsilon_0$ so that for $\epsilon \in (-\epsilon_1,\epsilon_1)$, $\gamma_\epsilon$ is nondegenerate.  By \cref{ChrisLemma}, $\epsilon_1$ can be taken to depend lower-semicontinuously on $(v,q)$.  Thus it has a minimum value $\epsilon_2$ on the compact set $\{(v,q)\in \Pp \st q\in \overline{\Oo_0}\}$.

Then $\epsilon_2 > 0$ satisfies the following: for each $\epsilon \in (-\epsilon_2,\epsilon_2)$ and for each $(q_0,q_1) \in \Oo_0 \times \Oo_0$, we have chosen a nondegenerate duration-$1$ solution to \cref{EOMepsilon} connecting $q_0$ to $q_1$, and the set of all these chosen paths for a given $\epsilon$ is precisely the set of solutions $\gamma_\epsilon$ to \cref{EOMepsilon} with $\bigl(\dot\gamma_\epsilon(0),\gamma_\epsilon(0)\bigr) \in \Pp$ and $\gamma(0),\gamma(1) \in \Oo_0$.  Thus this family $\gamma_{[-]}$ is smooth.

However, if $\epsilon > 0$ and $\gamma_\epsilon$ is a duration-$1$ solution to \cref{EOMepsilon}, then $\tau \mapsto \gamma_\epsilon(\epsilon^{-1}\tau)$ is a duration-$\epsilon$ solution to \cref{EOM} (\cref{EOMepsilon} with $\epsilon=1$).  In particular, we have constructed a smooth family $\gamma$ of classical paths with boundary values ranging in $\Oo_0 \times\Oo_0 \times (0,\epsilon_2)$.

Moreover, suppose that $\gamma$ is a duration-$\epsilon$ solution to \cref{EOM} with initial conditions $\gamma(0)\in \Oo_0$ and $\dot\gamma(0) = -B\bigl(\gamma(0)\bigr)$.  Then if $\epsilon$ is sufficiently small depending on $\gamma(0)$, we have $\bigl(\epsilon\dot\gamma(0), \gamma(0)\bigr) \in \Pp$, and so $\gamma$ is a member of our family.  In particular, our family contains all paths with $\dot\gamma(0) = -B\bigl(\gamma(0)\bigr)$ that start and end in $\Oo_0$ with sufficiently small duration depending on the endpoints.

\subsection{The limit as \texorpdfstring{$t\to 0$}{t goes to 0}} \label{step3}

In \cref{step1}, we showed that the only contributions to an integral of the form 
\begin{multline} \label{integraltodo}
\int^\formal_K \exp\bigl((i\hbar)^{-1}V_\gamma(t,q_0,q_1)\bigr)\,g(q_0)\,dq_0  = \\
= \int^\formal_K e^{-(i\hbar)^{-1}S_\gamma(t,q_0,q_1)} \sqrt{\left| \det \frac{\partial^2[-S_\gamma]}{\partial q_0\partial q_1}\right|} \exp\bigl({\textstyle \sum}\text{diagrams}\bigr)\,g(q_0)\,dq_0
\end{multline}
come from families $\gamma$ that include paths with initial conditions  $\dot\gamma(0) = -B\bigl(\gamma(0)\bigr)$.  Indeed, for fixed $t,q_1$, \cref{integraltodo} is supported only at the points $q_0 \in K$ so that the duration-$t$ path with initial conditions $\bigl(\dot\gamma(0),\gamma(0)\bigr) = \bigl(-B(q_0),q_0\bigr)$ ends at $\gamma(t) = q_1$.  In \cref{step2} we showed that for $\Oo = (0,\epsilon_2) \times \Oo_0 \times \Oo_0$ there is a unique such family, and that for $q_1 \in \Oo_0$ and sufficiently small $t$ depending on $q_1$, there is a unique such point $q_0$.  Indeed, we have 
 $q_1 - q_0 = -t\,B(q_0) + O(t^2)$.  We remark also that $\exp\bigl(\sum\text{diagrams}\bigr) =1 +  O(\hbar)$.

We claim that for these triples $(t,q_0,q_1)$, we have:
  \begin{equation} \label{Sesteqn}
    \Biggl( \! \biggl( \frac{\partial^2 [-S_\gamma]}{\partial q_1\partial q_0}\biggr)^{\!-1}\Biggr)^{\!kl} = t\,\delta^{kl} + O(t^2)
  \end{equation}
Indeed, consider integrating \cref{EOM} twice, with initial conditions $(\dot\gamma(0),\gamma(0)) = (v_0,q_0)$:
\[ \gamma^i(t) = q_0^i + tv_0^i + \int_0^t \int_0^\tau \left.\left( \left(\frac{\partial B_k}{\partial q^i} - \frac{\partial B_i}{\partial q^k}\right)\dot\gamma^k(\varsigma) - \frac{\partial C}{\partial q^i}\right)\right|_{q = \gamma(\varsigma)}\,d\varsigma\,dt\]
Then holding $q_0$ fixed we have:
\begin{multline} \label{Sesteqn1}
\frac{\partial \gamma^i(t)}{\partial v_0^j} = t\,\delta^i_j + \int_0^t \int_0^\tau \left.\left( \frac{\partial}{\partial q^l} \left(\frac{\partial B_k}{\partial q^i} - \frac{\partial B_i}{\partial q^k}\right)\frac{\partial \gamma^l(\varsigma)}{\partial v_0^j}\,\dot\gamma^k(\varsigma) + \mbox{} \right.\right.\\ \left. \left.\mbox{}
+ \left(\frac{\partial B_k}{\partial q^i} - \frac{\partial B_i}{\partial q^k}\right) \frac{\partial \dot\gamma^k(\varsigma)}{\partial v_0^j} -  \frac{\partial}{\partial q^l}\left( \frac{\partial C}{\partial q^i}\right)\right)\right|_{q = \gamma(\varsigma)}\,d\varsigma\,dt\end{multline}
We now take $t$ to be sufficiently small and $v_0$ to vary only in a compact neighborhood of $-B(q_0)$.  Then $\frac{\partial \gamma(\varsigma)}{\partial v_0}, \frac{\partial\dot\gamma(\varsigma)}{\partial v_0} = O(1)$ as $t\to 0$, but then \cref{Sesteqn1} gives $\frac{\partial \gamma^i(t)}{\partial v_0^j} = t\,\delta^i_j + O(t^2)$.  On the other hand, as we observed in \cref{Sesteqn2}, it follows from \cref{wellknown2} that $ \bigl( \frac{\partial^2 [-S_\gamma]}{\partial q_1\partial q_0}\bigr)^{-1} = \bigl( \frac{\partial v_0}{\partial q_1}\bigr)^{-1} = \frac{\partial q_1}{\partial v_0} $.  This gives \cref{Sesteqn}.

We make a similar argument, now holding $q_1$ fixed:
\begin{gather}
  q_0^i  = q_1^i - tv_0^i - \int_0^t \int_0^\tau \left.\left( \left(\frac{\partial B_k}{\partial q^i} - \frac{\partial B_i}{\partial q^k}\right)\dot\gamma^k(\varsigma) - \frac{\partial C}{\partial q^i}\right)\right|_{q = \gamma(\varsigma)}\,d\varsigma\,dt \\
  \frac{\partial q_0^i}{\partial v_0^j}  = -t\delta^i_j + O(t^2) \\
  \Biggl( \! \biggl( \frac{\partial^2 [S_\gamma]}{(\partial q_0)^2}\biggr)^{\!-1}\Biggr)^{\!kl}  = \Biggl( \! \biggl( -\frac{\partial}{\partial q_0} \bigl( v_0 + B(q_0)\bigr) \biggr)^{\!-1}\Biggr)^{\!kl} = \Bigl(\bigl( t^{-1}\delta + O(1)\bigr)^{-1}\Bigr)^{kl} = t\delta^{kl} + O(t^2)
\end{gather}

Then, by \cite[Theorem 3.15(ii)]{EZ2007}, we have:
\begin{multline*}
  \int^\formal \exp\bigl((i\hbar)^{-1}V_\gamma(t,q_0,q_1)\bigr)\,g(q_0)\,dq_0 = \\
  =  (2\pi \hbar)^{n/2} \times e^{i n \pi / 4} \times \det \bigl(t\delta^i_j + O(t^2)\bigr)^{-1/2} \times e^{ -(i\hbar)^{-1}S_\gamma(t,q_1 + B(q_1) + O(t^2),q_1)}\times \mbox{} \\
  \mbox{} \times \det\bigl( t\delta^i_j+ O(t^2)\bigr)^{1/2}\times g\bigl(q_1 + B(q_1) + O(t^2)\bigr)\times \bigl(1 + O(\hbar)\bigr)
\end{multline*}
But $S_\gamma(t,q_1+B(q_1) + O(t^2),q_1) = \int_0^t \bigl( \frac12 |\dot\gamma|^2 + B(\gamma)\cdot \dot\gamma - C(\gamma)\bigr)\,d\tau = \int_0^t \bigl( \frac12 (\dot\gamma + B)^2 + O(1)\bigr)\,d\tau = \int_0^t \bigl(O(t) + O(1)\bigr)d\tau = O(t)$.
Thus:
\begin{equation} \label{anotherintegral}
   \lim_{t\to 0} \int^\formal \exp\bigl((i\hbar)^{-1}V_\gamma(t,q_0,q_1)\bigr)\,g(q_0)\,dq_0 =(2\pi \hbar)^{n/2} e^{i n \pi / 4} \bigl( g(q_1) + O(\hbar)\bigr)
\end{equation}

What about the order-$\hbar$ terms?  They can be given explicitly as a sum along the lines of our \cref{Vdefine} --- c.f.\ \cite{Polyak2005,KolyaLong,melong} --- or implicitly as in \cite{EZ2007}.  All we need is the following fact: each degree in $\hbar$ is given by a finite sum in which for each summand there is some $M$ so that the summand scales as $\bigl( \frac{\partial^2 [-S_\gamma]}{(\partial q_0)^2}\bigr)^{-M}$ times a product of strictly fewer than $M$ terms given by derivatives of $g$ and $V_\gamma$ at $q_0 = q_1 + O(t^2)$.

The calculations in \cref{MainThmProof} in fact allow us to write down all derivatives of $V_\gamma$ with respect to $q_0$.  In addition to the derivatives computed in \cref{cancellations}, we have:
\begin{equation} \label{anewsum}
  \frac{\partial^k V_\gamma}{(\partial q_0)^k} = \sum_{\substack{\text{connected marked} \\ \text{Feynman diagrams $(\Gamma,M)$} \\ \text{with $n$ ordered  \tikz[baseline=(gamma.base)] \path node[inner sep=0pt] (gamma) {$\scriptstyle \gamma$} node[draw,dotted,thick,circle,inner sep=1pt,fit=(gamma)] (C) {} (C.north) ++(1pt,2pt) node[anchor= east] {$\scriptscriptstyle q_0$};s}}} \frac{(i\hbar)^{\lambda(\Gamma)}\Ff_\gamma(\Gamma,M)}{\left|\Aut(\Gamma,M)\right|}, \quad k\geq 3
\end{equation}
Consider expanding each edge as in \cref{greeneqn2}.  Then in each diagram a marked half edge converts some $\frac{\partial \gamma}{\partial q_a}$ into $\frac{\partial \dot\gamma}{\partial q_a} = -B(q_0) + O(t)$.  On the other hand, each vertex corresponds to an integration $\int_0^t$ of bounded functions, and so the vertices are $O(t)$.  Thus, by \cref{Sesteqn}, $\Ff_\gamma(\Gamma,M) = O(t^{|V|+|E|})$ for each diagram $\Gamma$ in \cref{anewsum}.  As the number of diagrams at each order is finite, this proves that for $k\geq 3$, $\frac{\partial^k V_\gamma}{\partial q_0^k} = O(t)$.

But the derivatives of $g$ do not depend on $t$; $\frac{\partial V_\gamma}{\partial q_0} = -\frac{\partial S_\gamma}{\partial q_0} + \sum\{\text{diagrams}\} = 0 + \sum\{\text{diagrams}\} = O(t)$; and $\frac{\partial^2 V_\gamma}{(\partial q_0)^2} = -\frac{\partial^2 S_\gamma}{(\partial q_0)^2} + \sum\{\text{diagrams}\} = O(t^{-1})$.  As each term after the first one in the asymptotic expansion of \cref{anotherintegral} has strictly more occurrences of $\bigl( \frac{\partial^2 [-S_\gamma]}{(\partial q_0)^2}\bigr)^{-1}$ than of derivatives of $g$ and $V_\gamma$, we see that the $O(\hbar)$ part of \cref{anotherintegral} is also $O(t)$.  Thus, as formal expressions in $\hbar$:
\begin{equation}
   \lim_{t\to 0} \int^\formal \exp\bigl((i\hbar)^{-1}V_\gamma(t,q_0,q_1)\bigr)\,g(q_0)\,dq_0 =(2\pi \hbar)^{n/2} e^{i n \pi / 4}g(q_1)
\end{equation}
\Cref{IVPthm} is proved.

%\bibliography{Edited}{}
%\bibliographystyle{plain}

%

\end{document}